\documentclass[fleqn,usenatbib,useAMS]{mnras}
\usepackage[utf8]{inputenc}
\usepackage{amsmath,amssymb,amsfonts}
\usepackage{mathptmx}
\usepackage[T1]{fontenc}
\usepackage[dvips]{graphicx}
\usepackage{subcaption}
\usepackage{amssymb}
\usepackage{arydshln}
\usepackage{natbib}
\usepackage{multirow}
\usepackage{ulem}
\usepackage{xcolor}
\usepackage{orcidlink}

\DeclareRobustCommand{\DEV}[3]{#3} 

\title[Eruptive novae in symbiotic systems]{Eruptive novae in symbiotic systems}
\author[I. B. Vathachira et al.]{
 Irin Babu Vathachira$^{1\, \orcidlink{0009-0003-3561-5961}}$, Yael Hillman  $^{2,1 \, \orcidlink{0000-0002-0023-0485}}$and Amit Kashi$^{1,3\, \orcidlink{0000-0002-7840-0181}}$\thanks {Amit Kashi e-mail: \url{kashi@ariel.ac.il}}\\
	$^{1}$Department of Physics, Ariel University, Ariel, 4070000, Israel\\
       $^{2}$Department of Physics, Technion -- Israel Institute of Technology, Haifa 3200003, Israel\\
    $^{3}$Astrophysics, Geophysics and Space Science (AGASS) Center, Ariel University, Ariel, 4070000, Israel\\
}

\date{Accepted 2023 November 13. Received 2023 November 8; in original form 2023 August 23}

\pubyear{2023}
\begin{document}
\maketitle
\begin{abstract}
      We conduct numerical simulations of multiple nova eruptions in detached, widely separated symbiotic systems that include an asymptotic giant branch (AGB) companion to investigate the impact of white dwarf (WD) mass and binary separation on the evolution of the system. The accretion rate is determined using the Bondi-Hoyle-Lyttleton method, incorporating orbital momentum loss caused by factors such as gravitational radiation, magnetic braking, and drag. The WD in such a system accretes matter coming from the strong wind of an AGB companion until it finishes shedding its envelope. This occurs on an evolutionary time scale of $\approx 3 \times 10^5$ years. Throughout all simulations, we use a consistent AGB model with an initial mass of $1.0 \mathrm {M_\odot}$ while varying the WD mass and binary separation, as they are the critical factors influencing nova eruption behavior. We find that the accretion rate fluctuates between high and low rates during the evolutionary period, significantly impacted by the AGB's mass loss rate. We show that unlike novae in cataclysmic variables, the orbital period may either increase or decrease during evolution, depending on the model, while the separation consistently decreases. Furthermore, we have identified cases in which the WDs produce weak, non-ejective novae and experience mass gain. This suggests that provided the accretion efficiency can be achieved by a more massive WD and maintained for long enough, they could potentially serve as progenitors for type Ia supernovae.
\end{abstract}
\section{Introduction}

Novae are bright outbursts that take place on the surface of white dwarfs (WDs) in binary systems. They were once referred to as 'guest stars', due to their sudden appearance in the sky. The WD accretes hydrogen rich matter from its donor that piles up onto the outer surface of the WD, slowly raising the sub-surface pressure and temperature \cite[]{1978ARA&A..16..171G,2011arXiv1111.4941B}. When the accreted mass reaches a critical amount --- sufficient to raise the temperature to the hydrogen burning limit --- the hydrogen is ignited \cite[]{1987Ap&SS.131..379S,2016PASP..128e1001S}. Since this occurs under degenerate conditions, the energy released serves to raise the temperature further, which in turn increases the burning rate, resulting in a runaway burning process --- a thermonuclear runaway (TNR) followed by the ejection of the envelope--- the nova \cite[]{1971MNRAS.152..307S,1972ApJ...176..169S,1978A&A....62..339P,1978ApJ...226..186S,1986ApJ...310..222P,2016PASP..128e1001S,2022ApJ...938...31S}. The concept of envelope ejection appeared as a momentary outburst in the works of \cite{2011A&A...533L...8S, 2013A&A...553A.123S, 2016A&A...590A.123S}, and \cite{2018ApJ...853...27M, 2020A&A...635A.115M}, while \cite{1943POMic...8..149M,1947PASP...59..244M}, \cite{1936PASP...48...29R}, and \cite{1987A&A...180..155F} argued that there would be multiple stages of outbursts. Spectroscopic studies of a dozen Galactic novae by \cite{2020ApJ...905...62A} have shown that the ejected mass is not ejected all at once but rather in a few epochs, culminating as mass flows of different velocities. Multiple flows of mass at different velocities may result in powerful interaction between them, resulting in shocks that may produce multi-band radiation \cite[]{2020NatAs...4..776A}. The expanding mass shell causes the light curve to rise to a maximum value with a brightness of a few $\sim 10^4\mathrm{ L_\odot}$ times the solar luminosity, which then decays within days or months \cite[]{1964gano.book.....P,1978ARA&A..16..171G}.

Novae may either occur in Cataclysmic Variables (CVs) or in symbiotic systems. The former comprises a red dwarf (RD) companion overfilling its Roche-lobe; thus, mass will transfer from it, through the inner Lagrangian point $L_1$, and accumulate on the surface of the WD \cite[]{1941ApJ....93..133K,1971ARA&A...9..183P,1987Ap&SS.131..379S}. In CVs the characteristics of novae are mainly determined by the mass of the WD and its accretion rate \cite[]{1978ApJ...222..604P,1982ApJ...257..312P,1984ApJ...281..367P,1994ApJ...424..319K,1995ApJ...445..789P,2005ApJ...623..398Y,2012BaltA..21...76S}. The latter is a system that hosts a red giant (RG) or a star on the asymptotic giant branch (AGB) that is losing mass via stellar winds \cite[]{1983ApJ...273..280K,2021MNRAS.501..201H}. Some of the wind can be gravitationally captured by the WD, accumulate, and eventually lead to a nova eruption.

Most simulations of nova evolution over the past few decades have disregarded binary interaction by assuming a constant accretion rate and obtained an understanding of the behavior of a nova for any given combination of WD mass and accretion rate (as well as some other parameters that have a secondary influence, such as the core temperature) \cite[]{1972ApJ...176..169S,1984ApJ...281..367P,1992ApJ...394..305K,1995ApJ...445..789P,2005ApJ...623..398Y,2007MNRAS.374.1449E,2013MNRAS.433.2884I,2013ApJ...777..136W,2015MNRAS.446.1924H}. Using the combined evolution code for self-consistent simulations of both binary components, \cite{2020NatAs...4..886H} allowed, for the first time, for the accretion rate to be determined organically at each numerical timestep and found that the accretion rate is highly sensitive to the amount of Roche-lobe overflow (RLOF), thus leading to rates that are not at all constant, but change --- decreasing and increasing over a single accretion phase, as well as having a secular decrease of many orders of magnitude.

For novae in widely separated symbiotic systems, RLOF does not occur (or is negligible), but the separation between the stellar components still plays an important role in determining the accretion rate, as does the wind rate from the giant \cite[]{2021MNRAS.501..201H}. Evolved giants (i.e., AGBs) are characterised by strong stellar winds due to radiation pressure on the outer layers of their atmosphere \cite[]{1983A&A...127...73B,1993ApJ...413..641V,dorfi1998agb}. 
Stars on the AGB experience wind patterns that can increase and decrease repeatedly by a factor of 10 - 100 due to thermal pulses in the envelope. These thermal pulses occur due to alternating burning in their outer helium and hydrogen shells \cite[]{1975ApJ...196..525I,1978PASJ...30..467S,doi:10.1146/annurev.aa.21.090183.001415}. Burning in the helium shell exerts radiation pressure outwards, causing the hydrogen shell to expand, thus, it cools, and the burning of hydrogen to helium halts. The helium continues to burn until the shell mass is reduced enough to substantially lower the fusion rate of helium, thus the radiation outwards is reduced, allowing the hydrogen shell to contract, heat, and begin fusing into helium again. This continues and thickens the helium layer again, until it becomes dense and hot enough to begin fusing again, and so the cycle continues until neither shell is capable of burning. These thermal pulses cause the radius of the AGB to increase and decrease accordingly, causing substantial changes to the mass loss (wind) rate \cite[]{1983A&A...127...73B,dorfi1998agb}. The mass loss rate for AGB stars is typically in the range  $10^{-7}-10^{-4}\mathrm{\mathrm{M_\odot yr^{-1}}}$ \cite[]{2009ASPC..414....3H}. These rates are influenced by factors such as thermal pulses, envelope expansion and contraction, stellar mass, luminosity, and metallicity.

In a symbiotic system, the orbital period is typically of the order of several years \cite[]{1999A&AS..137..473M,2019CoSka..49..189S}, although there are symbiotic systems with longer orbital periods, such as V407 Cyg with an orbital period of $\sim43$ years \cite[]{1990MNRAS.242..653M,8f7283202fed4be0a6cdf1aa0ef8c534}. This system has a very wide separation, of order  $10^3$$\mathrm{R_{\odot}}$ meaning that any mass that might possibly pass through $L_1$ (i.e., via RLOF) is only a negligible fraction of the mass that may be captured via Bondi-Hoyle-Lyttleton (BHL) accretion \cite[]{1939PCPS...35..405H,1944MNRAS.104..273B,1952MNRAS.112..195B}. The BHL method was developed to demonstrate the manner in which a point star or object moving in a cloud of gas may accrete matter from the cloud. \cite{1939PCPS...35..405H} considered a point mass (star) moving in an ideal fluid at relatively high velocity (supersonic velocity). The gravity acting on the star (or point mass) creates a shock, and when the gas decelerates through the shock, it loses momentum and, thus velocity. Some of the gas becomes gravitationally bound to the star --- accreted. The basic principles of the BHL method may also be used to describe a WD in a cloud of wind coming from a giant companion as in symbiotic systems --- as described above. Thus, the wind is assumed to escape the AGB isotropically, and the fraction of it that will be captured by the WD is determined by the BHL prescription \cite[]{1939PCPS...35..405H,1944MNRAS.104..273B} while the rest is lost from the system, taking with it angular momentum, thus causing the separation to shrink.

In this study we present self-consistent long-term simulations of the evolution of symbiotic systems that produce nova eruptions using models of different WD masses accreting matter from the wind of their AGB donors. This work is the natural continuation of the work done by \cite{2021MNRAS.501..201H} in which they compared simulations of RLOF and wind accretion for the same WD model with donors of the same mass --- a red dwarf (RD) for the CV and an AGB for the wind model. Here we consider the same AGB model while expanding the study by varying the WD mass and binary separation in order to assess --- for the first time --- the influence of these two parameters (separation and WD mass) on the systems' evolutions. We follow, for all our models (a total of nine), the evolution of various parameters, such as the change in mass of the WD, AGB, and separation of the binary system, the change in accretion rate as well as the orbital period ($\rm P_{orb}$) etc. Our methods of calculation, code and models are described in detail in section \ref{sec:models}. We present the results of our simulation in section \ref{sec:results}, followed by a discussion in section \ref{sec:discussion}. We summarize our main conclusions in section \ref{sec:conclusions}. 

\section{Simulations and models}\label{sec:models}

\begin{table*}
    \begin{tabular}{l c c c c c c c c c c}
    \hline  
      Model No. &  1 & 2 & 3 & 4 & 5 & 6 & 7 & 8 & 9\\
    \hline
      $M_{WD}$[$\mathrm{M_\odot}$] & 1.25 & 1.25 & 1.25 & 1.00 & 1.00 & 1.00 & 0.70 & 0.70 & 0.70\\
      $a$ [$10^3\mathrm{\mathrm{R_\odot}}$] & 2.0 & 3.4 & 5.0 & 2.0 & 3.4 & 5.0 & 2.0 & 3.4 & 5.0\\
      Number of cycles & 8092 & 2203 & 904 & 1006 & 261 & 105 & 122 & 30 & 14\\
      $\Delta M_{WD}$[$\mathrm{M_\odot}$] & +3.1$\times10^{-3}$ & +4.2$\times10^{-4}$ & -4.8$\times10^{-5}$ & +3.3$\times10^{-3}$ & +3.9$\times10^{-4}$ & -7.6$\times10^{-6}$ & +2.5$\times10^{-3}$ & +2.3$\times10^{-4}$ & -5.1$\times10^{-6}$\\
      $\dot{\overline{M}}_{\mathrm{acc,max}}$[$\mathrm{\mathrm{M_\odot yr^{-1}}}$]& 1.34$\times10^{-7}$ & 6.17$\times10^{-8}$ & 3.25$\times10^{-8}$ & 9.86$\times10^{-8}$ & 4.26$\times10^{-8}$ & 2.18$\times10^{-8}$ & 5.63$\times10^{-8}$ & 2.27$\times10^{-8}$ & 1.08$\times10^{-8}$\\ 
      
      $\dot{\overline{M}}_{\mathrm{acc,min}}$[$\mathrm{\mathrm{M_\odot yr^{-1}}}$]& 3.18$\times10^{-9}$ & 1.43$\times10^{-9}$& 7.92$\times10^{-10}$ & 3.05$\times10^{-9}$ & 1.49$\times10^{-9}$ & 9.97$\times10^{-10}$ & 3.70$\times10^{-9}$& 2.35$\times10^{-9}$ & 1.69$\times10^{-9}$\\
 $\dot{\overline{M}}_{\mathrm{max}}$/$\dot{\overline{M}}_{\mathrm{min}}$& 42.1 & 43.1 & 41.0 & 32.3 & 28.6& 21.8 & 15.2 & 9.65 & 6.39\\
 $\dot{\overline{M}}_{\mathrm{w,acc,max}}$[$\mathrm{\mathrm{M_\odot yr^{-1}}}$]& 2.76$\times10^{-6}$ & 2.99$\times10^{-6}$ & 3.07$\times10^{-6}$ & 2.89$\times10^{-6}$ & 3.04$\times10^{-6}$ & 3.09$\times10^{-6}$ & 3.00$\times10^{-6}$ & 3.08$\times10^{-6}$ & 3.00$\times10^{-6}$\\
 
 $\dot{\overline{M}}_{\mathrm{w,acc,min}}$[$\mathrm{\mathrm{M_\odot yr^{-1}}}$]& 8.20$\times10^{-8}$ & 8.10$\times10^{-8}$ & 8.38$\times10^{-8}$ & 1.21$\times10^{-7}$ & 1.23$\times10^{-7}$ & 1.56$\times10^{-7}$& 2.43$\times10^{-7}$ & 3.60$\times10^{-7}$ & 5.06$\times10^{-7}$\\
 
 $m_{\mathrm{acc,max}}$[$\mathrm{M_\odot}$] & 6.32$\times10^{-6}$  &6.58$\times10^{-6}$& 6.49$\times10^{-6}$ &3.47$\times10^{-5}$ & 3.43$\times10^{-5}$ &3.11$\times10^{-5}$ & 1.31$\times10^{-4}$ &1.21$\times10^{-4}$ & 1.11$\times10^{-4}$\\ 
 $m_{\mathrm{acc,min}}$[$\mathrm{M_\odot}$] & 1.26$\times10^{-6}$  &2.17$\times10^{-6}$ & 3.21$\times10^{-6}$ &6.80$\times10^{-6}$ & 1.29$\times10^{-5}$ &1.86$\times10^{-5}$ & 2.81$\times10^{-5}$ &6.78$\times10^{-5}$ & 6.53$\times10^{-5}$\\
      
        \hline
    \end{tabular}
    \caption{The rows are, from top to bottom: model number; initial WD mas ($M_{WD}$); initial separation ($a$); total number of cycles; change in WD mass ($\Delta M_{WD}$); maximum averaged cyclic accretion rate ($ \dot{\overline{M}}_{\mathrm{acc,max}}$); minimum averaged cyclic accretion rate ($\dot{\overline{M}}_{\mathrm{acc,min}}$); ratio of maximum to minimum averaged cyclic accretion rates ($\dot{\overline{M}}_{\mathrm{max}}$/$\dot{\overline{M}}_{\mathrm{min}}$);  averaged cyclic wind rate ($ \dot{\overline{M}}_{\mathrm{w,acc,max}}$); averaged cyclic wind rate ($\dot{\overline{M}}_{\mathrm{w,acc,min}}$) corresponding to maximum and minimum averaged cyclic accretion rate respectively; maximum accreted mass per cycle ($m_{\mathrm{acc,max}}$) and minimum accreted mass per cycle($m_{\mathrm{acc,min}}$).}
    \label{tab:Table 2}
\end{table*}

\subsection {The self-consistent code}\label{subsec:SCC}
We have considered a binary system comprising a WD accreting matter from the wind of its AGB donor.
We used the self-consistent, binary evolution code, as described in detail in \cite{2020NatAs...4..886H} which utilizes a hydro-static stellar evolution code \cite[]{2009MNRAS.395.1857K} and a nova evolution code \cite[]{1995ApJ...445..789P,2007MNRAS.374.1449E,2015MNRAS.446.1924H}. The former is capable of evolving a star from pre-MS, through the stellar evolution stages, until it loses its envelope, contracts and cools, eventually becoming a WD. The latter is a hydrodynamic Lagrangian nova evolution code used for simulating consecutive nova eruptions for a given initial WD mass and a constant accretion rate. In the combined code, \cite{2020NatAs...4..886H} use Roche geometry considerations following \cite{1983ApJ...268..368E}. Both the initial stellar masses (the WD's and the RD's) are provided, and the system is initially positioned in a manner that the RD fills its Roche lobe. As the system evolves over time, the accretion rate is determined organically at each time step, taking into account mass transfer from the RD to the WD, mass loss during each nova eruption, magnetic braking (MB), and gravitational radiation (GR) following \cite{2015ApJS..220...15P}. These factors contribute to changes in angular momentum, thereby affecting the separation between the stars, which in turn directly influences the accretion rate \cite[]{2020NatAs...4..886H,2021MNRAS.505.3260H}. Here, we use the modification from \cite{2021MNRAS.501..201H} where they adapted the code to simulate accretion from the wind of an AGB star in a widely separated system for which the accretion rate onto the surface of the WD is not due to RLOF but rather via the BHL method. Furthermore, in addition to angular momentum loss (AML) due to MB and GR, the engulfment of the WD in a cloud of wind induces a non-negligible drag force \cite[]{1976ApJ...204..879A,2021MNRAS.505.3260H} which was also taken into account during simulations. An AGB companion of mass 1.0$\mathrm{M_\odot}$ serves as our initial donor model for all the simulations explored in this work.

\subsection{Determining the accretion rate}\label{subsec:accretion rate}
We have made use of the Bondi-Hoyle-Lyttleton (BHL) accretion formula to calculate the accretion rate of wind from an AGB donor onto the surface of a WD. Following  \cite{1944MNRAS.104..273B}, equation \ref{1} gives the accretion radius ($r_a$) around the WD within which the material will be captured by the gravitational pull of the WD, and thus, be accreted onto its surface;

\begin{equation}\label{1}
    r_a=\frac{2GM_{WD}}{v^2_w},
\end{equation}
where $M_{WD}$ is the mass of WD, $v_w$ is the velocity of the wind relative to the WD, and $G$ is the gravitational constant. The accretion rate determined via the BHL method is given by equation \ref{2};
\begin{equation}\label{2}
    \dot{M}_{acc} = 4 \pi \rho_w \frac{G^2 M^2_{WD}}{(v^2_w + v^2_s)^{3/2}},
\end{equation}
where $\rho_w$ is the density of the AGB's wind at the location of the WD, and $v_s$ is the speed of sound in the cloud of gas. The density $\rho_w$ is given by equation \ref{3};
\begin{equation}\label{3}
    \rho_w= \frac{\dot{M}_{w}}{4 \pi a^2 v_{w}},
\end{equation}
where $\dot{M}_{w}$ is the mass loss rate (wind rate) \cite[]{1975ApJ...195..157C} and $a$ is the binary separation. We have adopted methods from the work of \cite{Kashi_2009} to include the orbital velocity ($v_{orb}$) of the WD and the terminal wind velocity ($v_\infty$) (typically $\sim$ 20 km $\mathrm{s^{-1}}$) \cite[]{Kashi_2009,2021MNRAS.501..201H}. The $v_{w}$ is given by equation \ref{4};
\begin{equation}\label{4}
    v_{w}= v_\infty+\frac{R_{AGB}}{a}(v_s-v_\infty),
\end{equation}
where $R_{AGB}$ is the radius of the AGB donor. The orbital velocity is given by equation \ref{5};
\begin{equation}\label{5}
   v^2_{\mathrm{orb}}=\frac{G(M_{WD}+M_{AGB})}{a},
\end{equation}
where $M_{AGB}$ is the mass of AGB. Combining Equations \ref{2} - \ref{5} gives the dependence of the accretion rate on the wind rate, wind velocity and orbital velocity as well as on the separation and the WD mass. Thus, the accretion rate can be calculated as
\begin{equation}\label{6}
    \dot{M}_{\mathrm{acc}} = \frac{G^2 M^2_{WD}}{a^2 v_{w}(v^2_{orb}+v^2_{w})^{3/2}} {\dot{M}_{w}}.
\end{equation}
 
\subsection{Angular Momentum Loss}\label{subsec:AML}
Though we have taken separations that are wide enough to ensure that the RLOF is negligible, the separation still has an effect on determining the amount of mass captured by the WD, as seen in equation \ref{6}.  Moreover, since an AGB star blows wind in all directions and only a small portion of it (determined by the BHL scheme) is captured by the WD, mass is lost from such a system at a much faster rate than in CVs \cite[]{1983A&A...127...73B,dorfi1998agb} for which we assume that all the mass that is pulled away from the donor piles up onto the WD. This has a considerable effect on AML and therefore, on the evolution of the separation.

Other factors that effect angular momentum in binary stars are MB and GR. The change in angular momentum due to MB and GR is given as \cite[]{2015ApJS..220...15P,2020NatAs...4..886H};
\begin{equation} \label{7}
    \dot{J}_{MB} = -1.06\times10^{20}M_{AGB}R^{4}_{AGB}P^{-3}_{\mathrm{orb}},\\ 
\end{equation}

\begin{equation} \label{8}
    \dot{J}_{GR} = - \frac{32}{5c^{5}}  \biggl(\frac{2\pi G}{P_{\mathrm{orb}}}\biggl)^{7/3} \frac{(M_{AGB}M_{WD})^{2}}{(M_{AGB}+M_{WD})^{2/3}},\\ 
\end{equation}
where $\dot{J}_{MB}$ and $\dot{J}_{GR}$ are the change in orbital angular momentum due to magnetic braking and gravitational radiation respectively and $c$ is the speed of light.

In symbiotic systems, where the WD is embedded in the wind of a giant, the drag force plays a crucial role. This force becomes a significant factor and removes angular momentum from the system by resisting the WD's motion as it moves through the wind emitted by its companion giant star. Friction and collision with particles in the wind reduce the orbital velocity (and kinetic energy) of the system \cite[]{1976ApJ...204..879A}. Thus, a WD experiencing drag force ($D_w$) undergoes orbital momentum loss and is given as;
\begin{equation} \label{9}
    D_w= \pi \rho_w r^2_a v^2_w,
\end{equation}
During the accretion phase of each cycle, the separation (and from it the accretion rate) is calculated at each time-step accounting for orbital momentum change due to MB, GR and drag, as described in Equation \ref{7}, \ref{8} and \ref{9} respectively. After each nova eruption, before resuming accretion, the separation (and accretion rate) is calculated as follows:
\begin{equation} \label{10}
     \Delta a = 2a \biggl(\frac{m_{\mathrm{ej}}-m_{\mathrm{acc}}}{M_{WD}} + \frac{\mathrm{m_{acc}}}{M_{AGB}}\biggl).
\end{equation}
to account for AML due to mass lost from the system, where $m_{\mathrm{acc}}$ and $m_{\mathrm{ej}}$ are the accreted and ejected masses of the preceding nova cycles. These methodologies follow the works of \cite{2020NatAs...4..886H,2021MNRAS.505.3260H} and \cite{2021MNRAS.501..201H}.

A more detailed elaboration of the calculations may be found in \cite{Kashi_2009}, \cite{2020NatAs...4..886H}, \cite{2021MNRAS.505.3260H} and \cite{2021MNRAS.501..201H}.

\subsection{Initial Model parameters} \label{subsec:Models}
We have carried out and analysed a total of nine different simulations --- three initial WD masses each used with three different initial binary separations. We used the same donor AGB model for all nine models --- a $1.0\mathrm{M_\odot}$ solar mass AGB star --- the donor model is obtained by following a 1.35$\mathrm{M_\odot}$ pre-main sequence star as it evolves through the main-sequence stage, becomes a red giant, goes through the horizontal branch stage, and finally emerges on the asymptotic giant branch (AGB), with a radius of about $177 \mathrm{\mathrm{R_\odot}}$ and a significant mass loss rate of order $\sim 10^{-6}-10^{-7}\mathrm{\mathrm{M_\odot yr^{-1}}}$. At this point the mass of the AGB is $\sim$ 1.0$\mathrm{M_\odot}$. Our WD models are for three different masses: $0.7\mathrm{M_\odot}$, $1.0\mathrm{M_\odot}$ and $1.25\mathrm{M_\odot}$ each used with three different separations: $2.0\times10^3\mathrm{\mathrm{R_\odot}}$, $3.4\times10^3\mathrm{\mathrm{R_\odot}}$ and $5.0\times10^3\mathrm{\mathrm{R_\odot}}$ as given in Table \ref{tab:Table 2}. We confined our simulations to wide separations to avoid the possibility of the donor filling its Roche-lobe \cite[]{2013A&A...552A..26A}. 
The simulations were carried out until the donor wind ceased, it has been eroded down to $\sim0.57\mathrm{M_\odot}$ and it began contracting to a WD. Since, we chose the same donor model for all the simulations, the total evolutionary time of all nine simulations is dictated by the donor's evolutionary time of  $\sim3\times10^5$yrs. 

\section{Results}\label{sec:results}

\begin{figure}
\begin{subfigure}[b]{0.5\textwidth}
   \includegraphics[trim={0.5cm 0.5cm 0.9cm 1.5cm},clip,width=1\columnwidth]{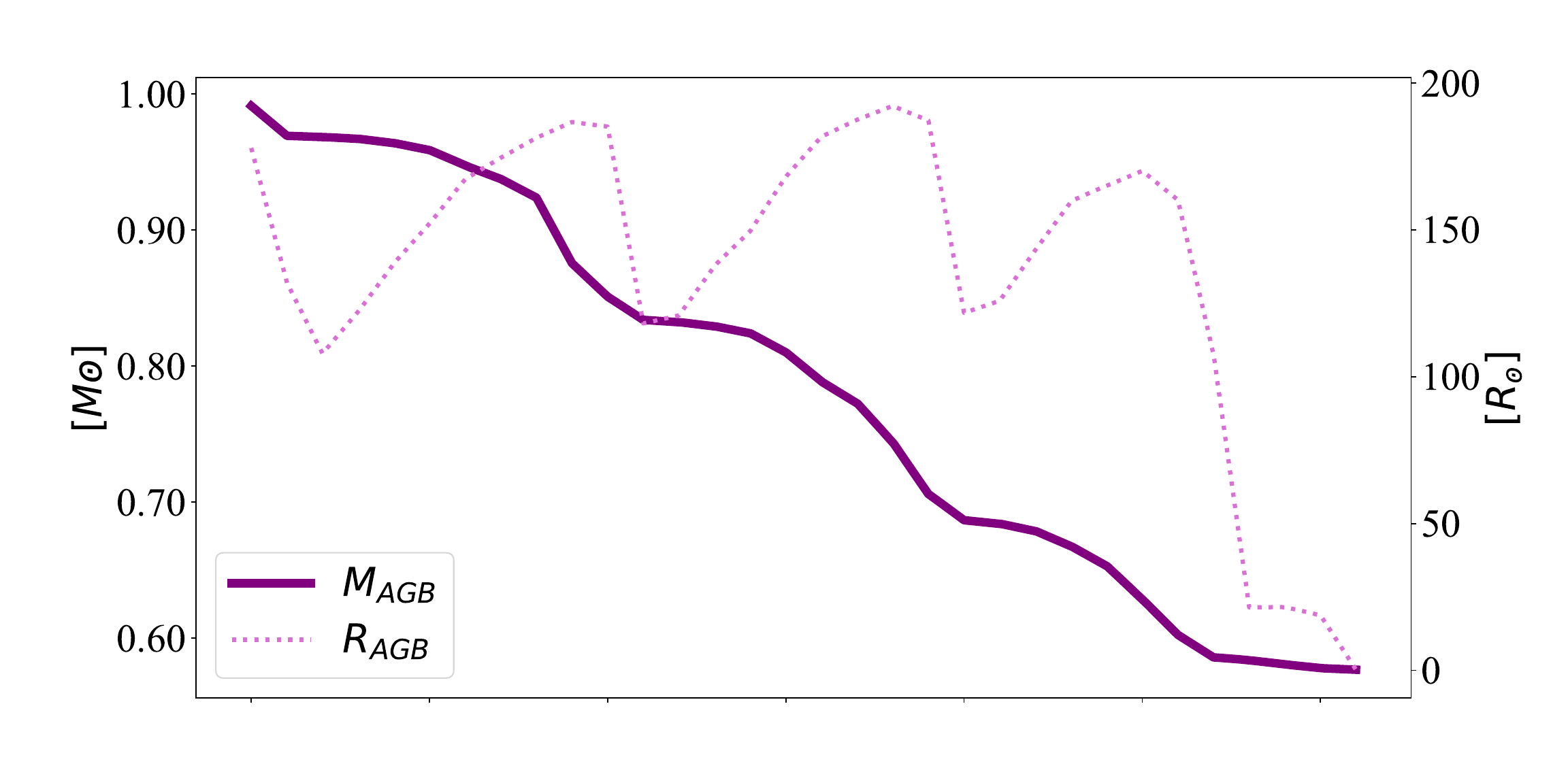}
\end{subfigure}
\begin{subfigure}[b]{0.5\textwidth}
   \includegraphics[trim={0.5cm 0.2cm 0.9cm 1.0cm},clip,width=1\columnwidth]{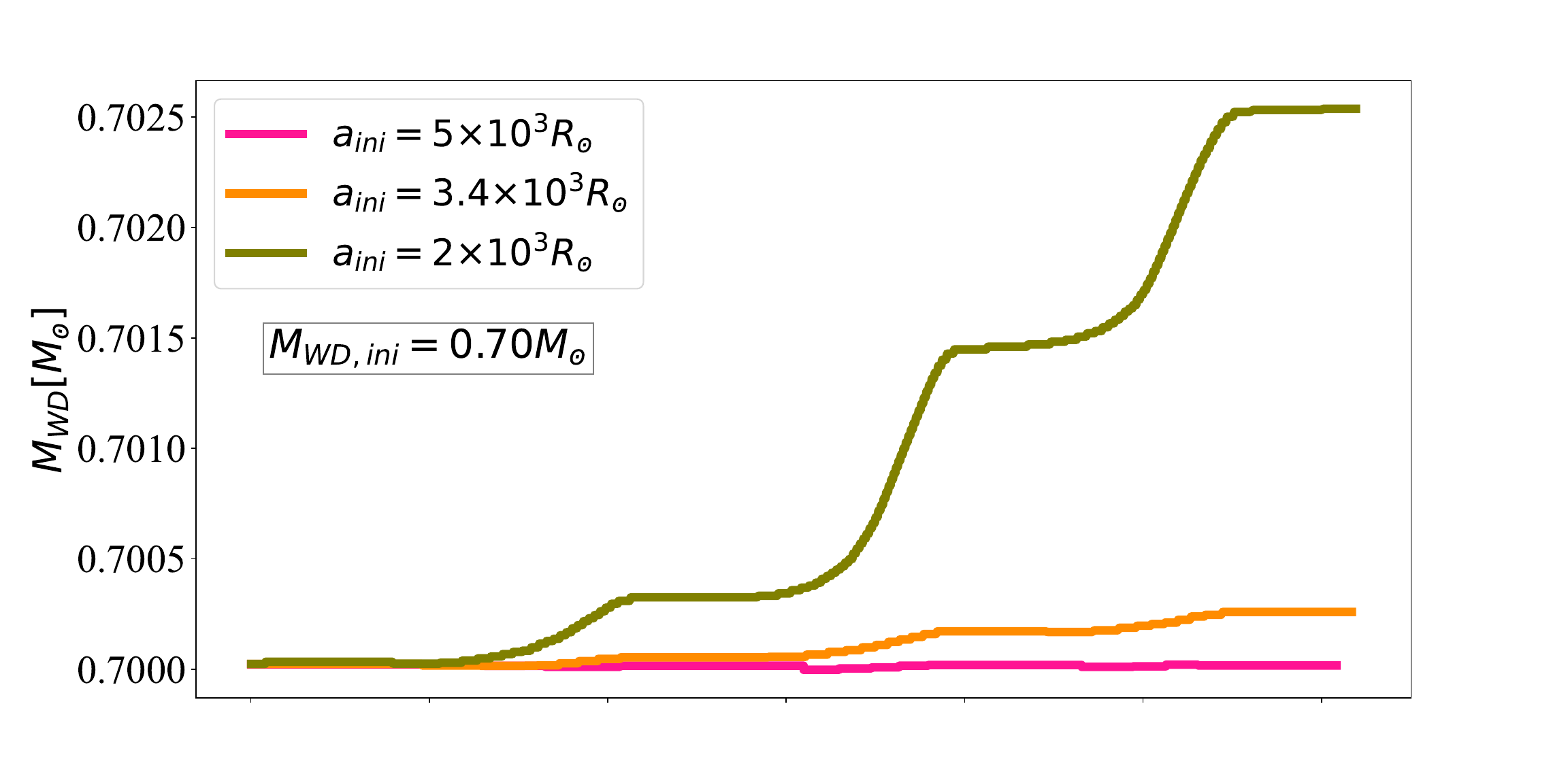}
\end{subfigure}
\begin{subfigure}[b]{0.5\textwidth}
   \includegraphics[trim={0.5cm 0.0cm 0.9cm 1.0cm},clip,width=1\columnwidth]{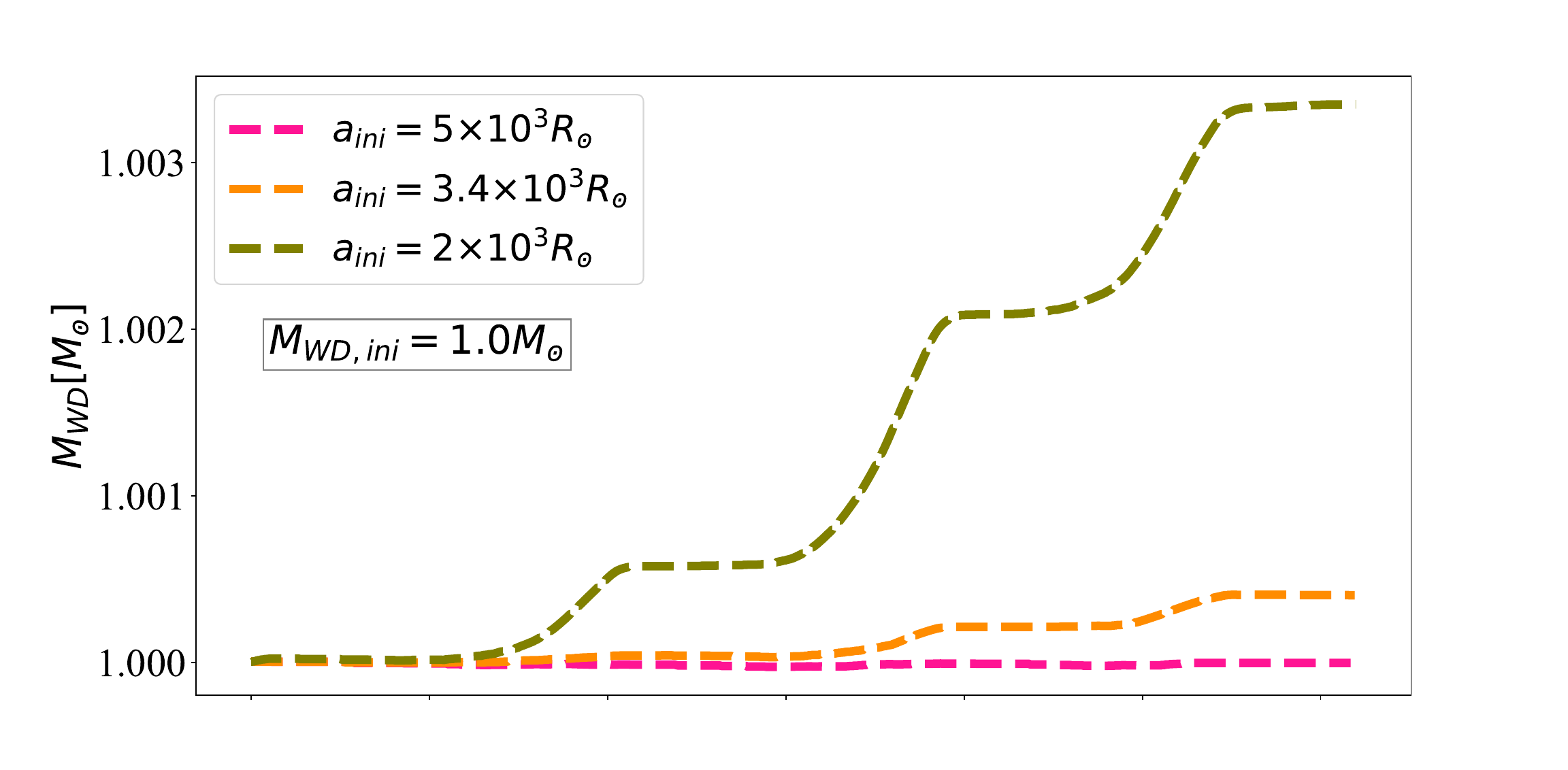}
\end{subfigure}
\begin{subfigure}[b]{0.5\textwidth}
   \includegraphics[trim={0.5cm 0cm 0.9cm 0.0cm},clip,width=1\columnwidth]{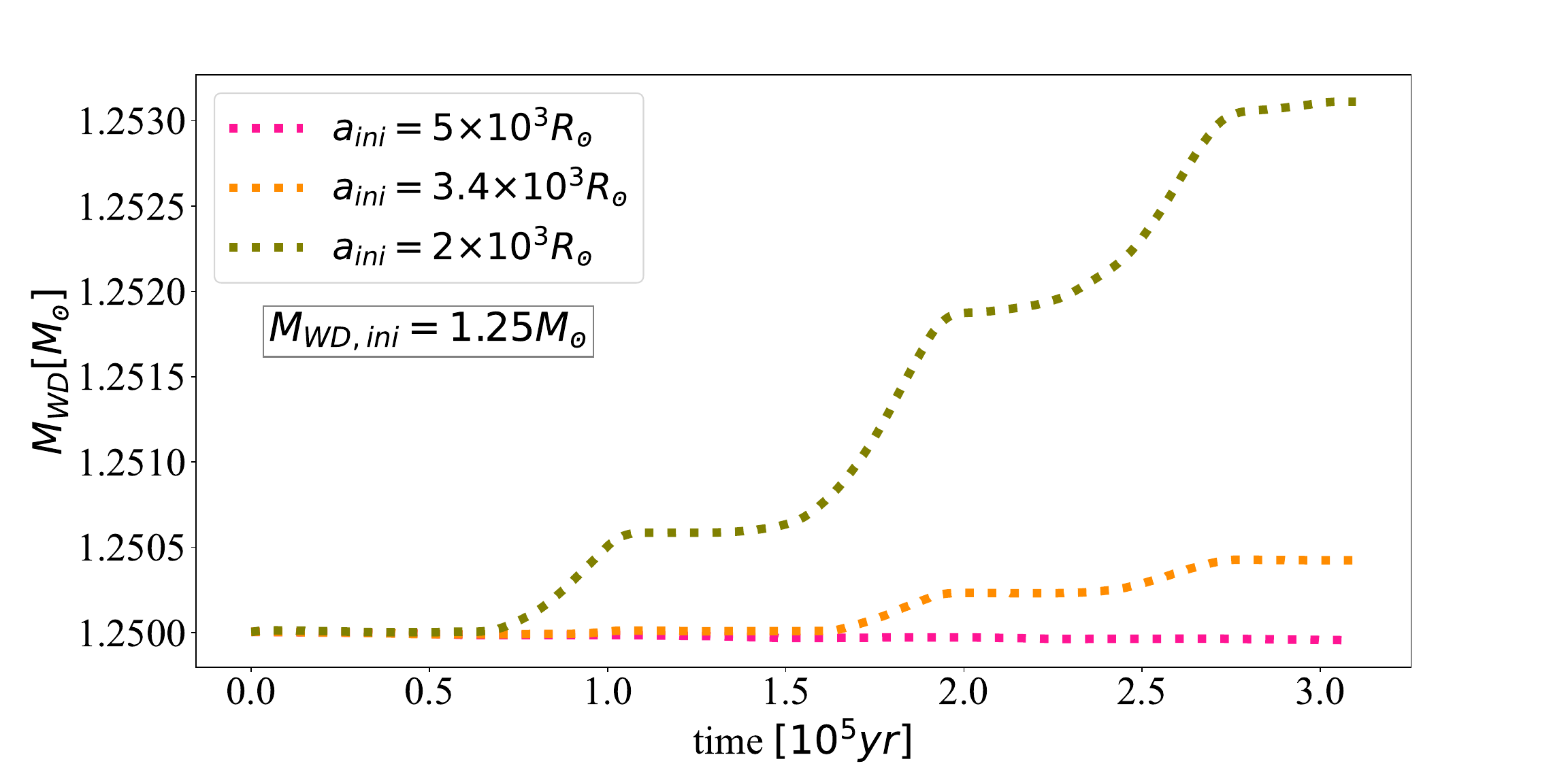}
\end{subfigure}

\caption{Mass-Radius evolution of the AGB (top) and the three WD masses for the three separations. For all the simulations the initial WD masses (from top to bottom) are marked as follows: $0.7 \mathrm{M_\odot}$- solid; $1.0 \mathrm{M_\odot}$- dashed; $1.25 \mathrm{M_\odot}$- dotted, and the initial separations are marked as: $2.0\times10^{3}\mathrm{\mathrm{R_\odot}}$- green; $3.4\times10^{3}\mathrm{\mathrm{R_\odot}}$- orange; $5.0\times10^{3}\mathrm{\mathrm{R_\odot}}$- pink. The most massive WD with the smallest separation experiences the highest mass gain, while those with the least separation experience mass loss.}\label{fig:Donor and WD}
\end{figure}

\begin{figure}
    \hspace*{-0.2cm}
    \includegraphics[trim={1.0cm 4cm 0.0cm 6.7cm},clip,width=1.1\columnwidth]{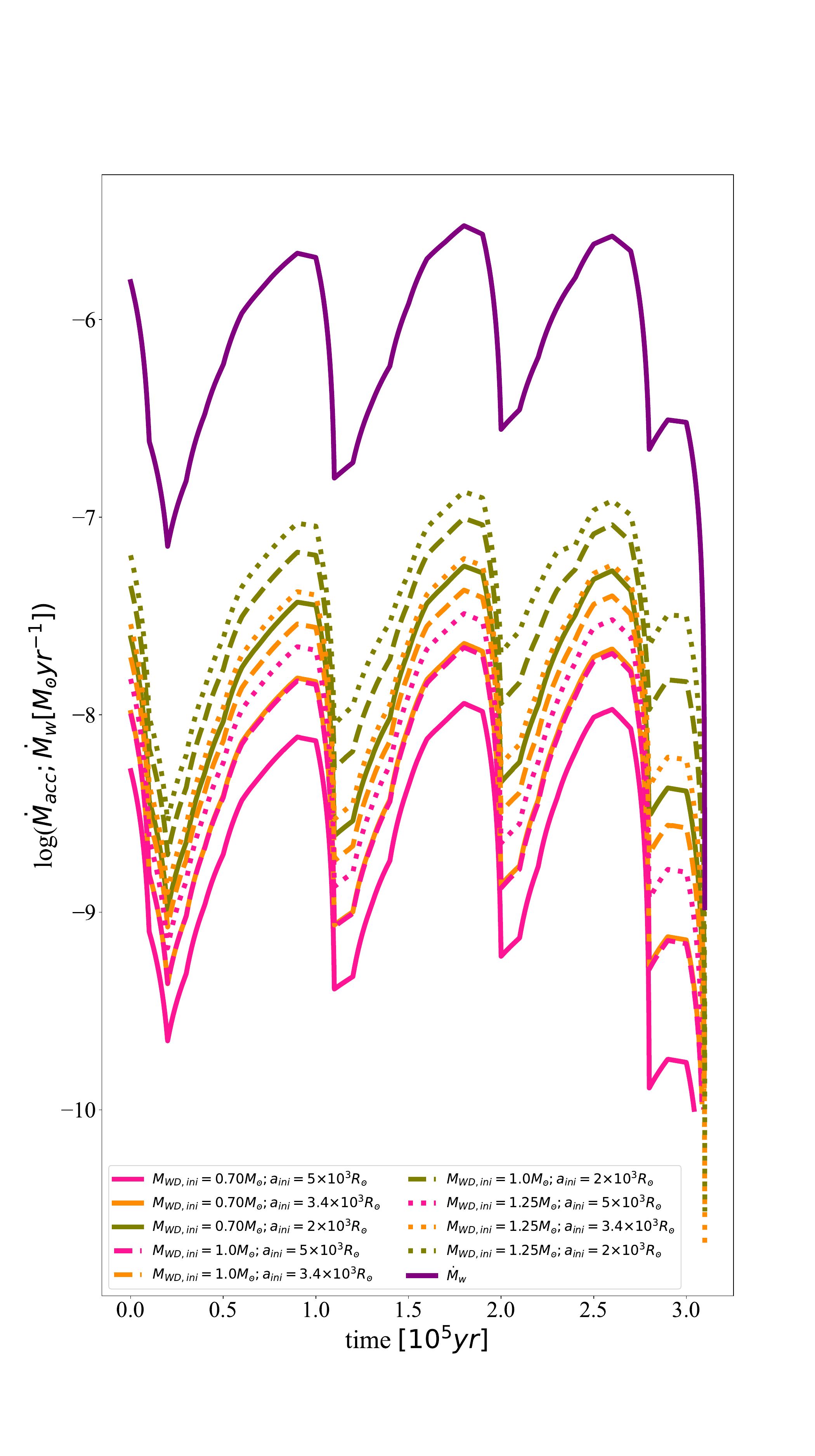}
    \caption{Evolution of the accretion rates ($\mathrm{\dot{M}_{acc}}$) for the nine models. Line type and color as described in Figure \ref{fig:Donor and WD}. The purple line shows the wind rate of the AGB ($\mathrm{\dot{M}_w}$). The fluctuations in the graph result from the surges in wind rate caused by thermal pulses, leading to periodic increases and decreases in the wind's intensity. The pattern of the accretion rate mirrors that of the wind rate. The decline in accretion rate towards the later stages occurs because, by that point, the donor has transformed into a WD, causing the wind rate to decrease substantially.}
    \label{fig:mdot}
\end{figure}

\begin{figure*}
    \hspace*{-2.5cm}
   \begin{subfigure}[b]{1\textwidth}
   \includegraphics[trim={0.4cm 0cm 0.4cm 0cm},clip ,width=1.15\columnwidth]{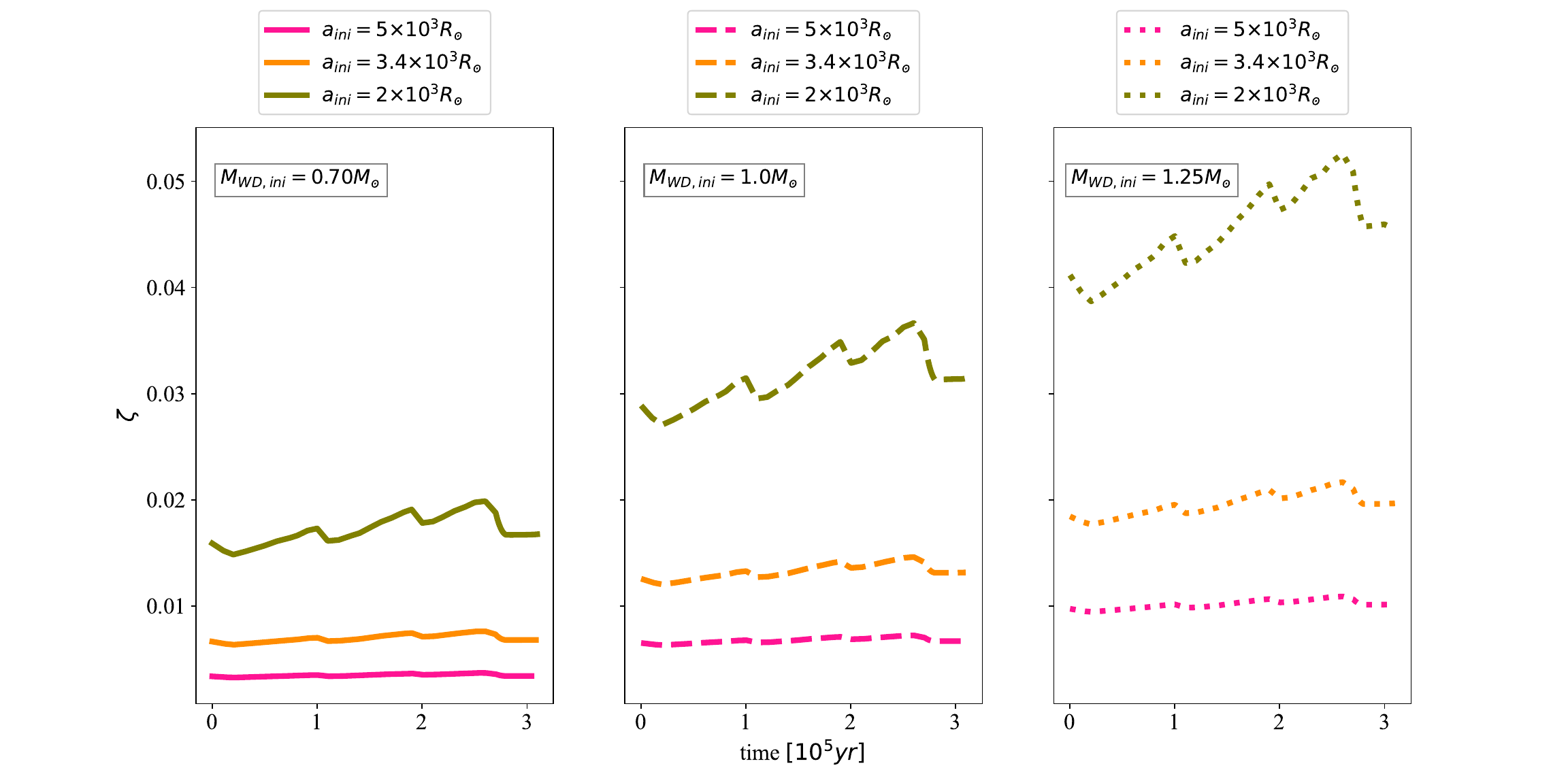}
   \end{subfigure}
   \caption{ Figure shows the mass \textit{transfer} efficiency, given as the ratio between the accretion rate and the wind rate ($\mathrm{\zeta}=\mathrm{\dot{M}_{acc}/\dot{M}_w}$), vs. time for the nine models. It illustrates the mass captured by the WD through the BHL mechanism. The parameter $\zeta$ grows with escalating white dwarf mass, and it diminishes as the separation widens. Line type and color as described in Figure \ref{fig:Donor and WD}.}
   \label{fig:Accretion rate}
\end{figure*}

\begin{figure*}
\hspace*{-1.3cm}
    \includegraphics[trim={2cm 4cm 2cm 4cm},clip ,width=2.4\columnwidth]{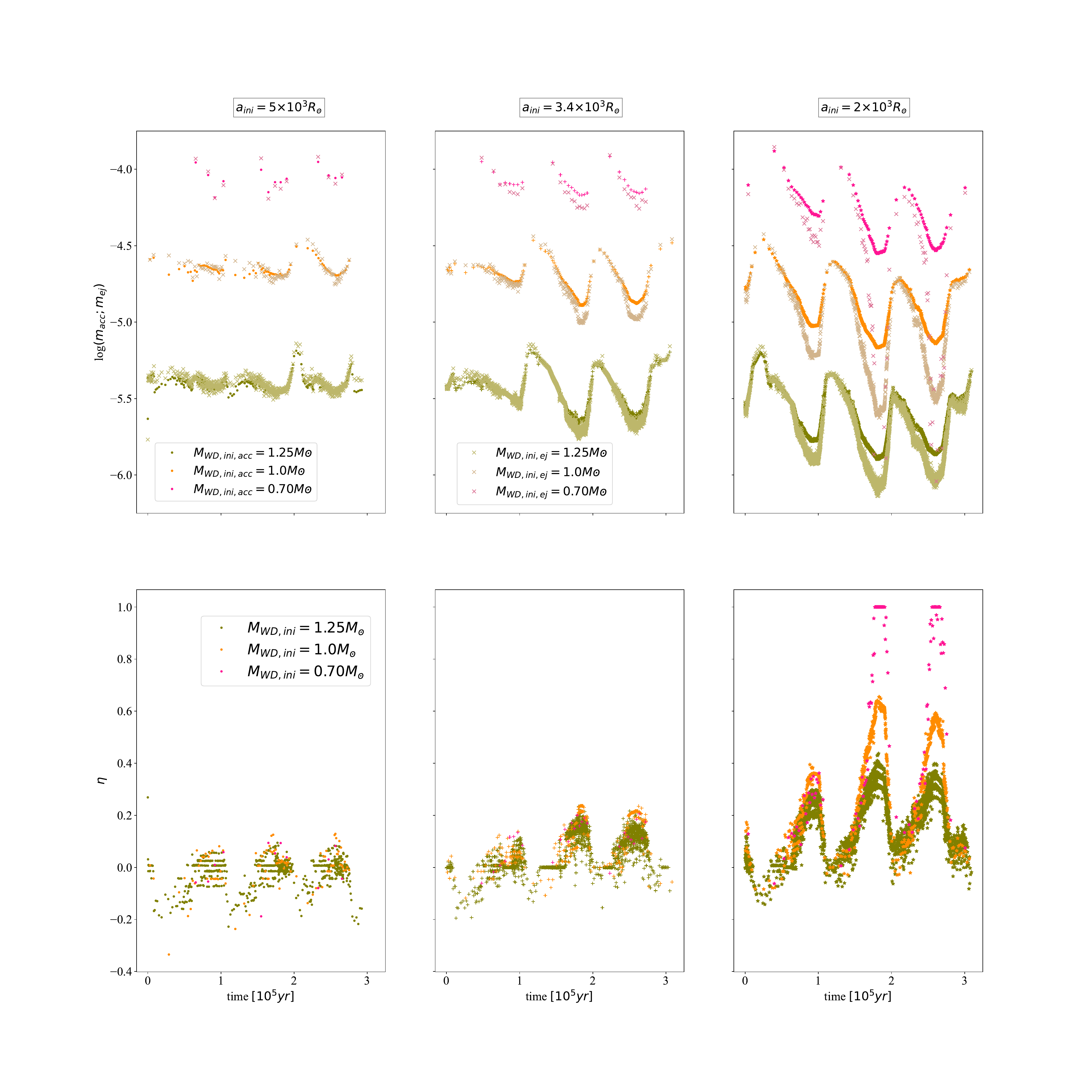}
   \caption{Figure shows the the amount of accreted and ejected matter on the surface of WD (top) and and the mass retention efficiency $\eta$ (bottom) vs. time. The color code of WD models are: green-$1.25\mathrm{M_\odot}$; orange-$1.0\mathrm{M_\odot}$ and pink-$0.70\mathrm{M_\odot}$. The first column represents when the separation is $a=5\times10^3 \mathrm{R_\odot}$; second column is $a=3.4\times10^3 \mathrm{R_\odot}$ and third column is $a=2\times10^3 \mathrm{R_\odot}$. Accreted masses exhibit an increase with decreasing WD mass, while wider separations correspond to higher ejected masses. This behavior stems from the inverse relationship between separation and accretion rates. The mass retention efficiency $\eta$, diminishes as the separation grows. The upper limit for $\eta$ is 1, for non-ejective cycles. The negative $\eta$ values are linked with reduced accretion rates.}
   \label{fig:mejtomacc}
\end{figure*}

\begin{figure}
\hspace*{-0.5cm}
\begin{subfigure}[b]{0.5\textwidth}
   \includegraphics[trim={0.0cm 0.0cm 0.0cm 0cm},clip,width=1.04\columnwidth]{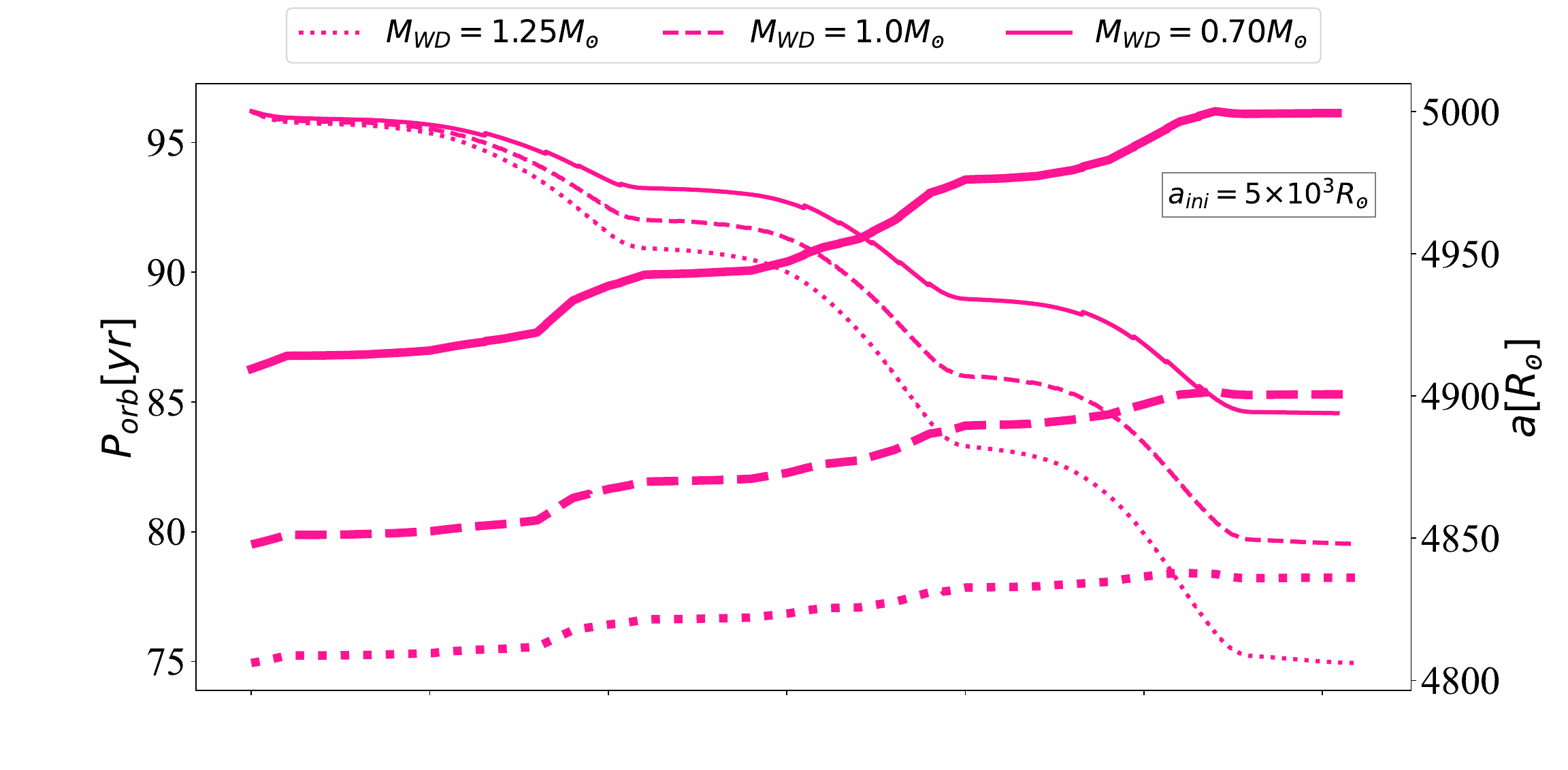}
\end{subfigure}
\hspace*{-0.5cm}
\begin{subfigure}[b]{0.5\textwidth}
   \includegraphics[trim={0.0cm 0.0cm 0.0cm 0.0cm},clip,width=1.04\columnwidth]{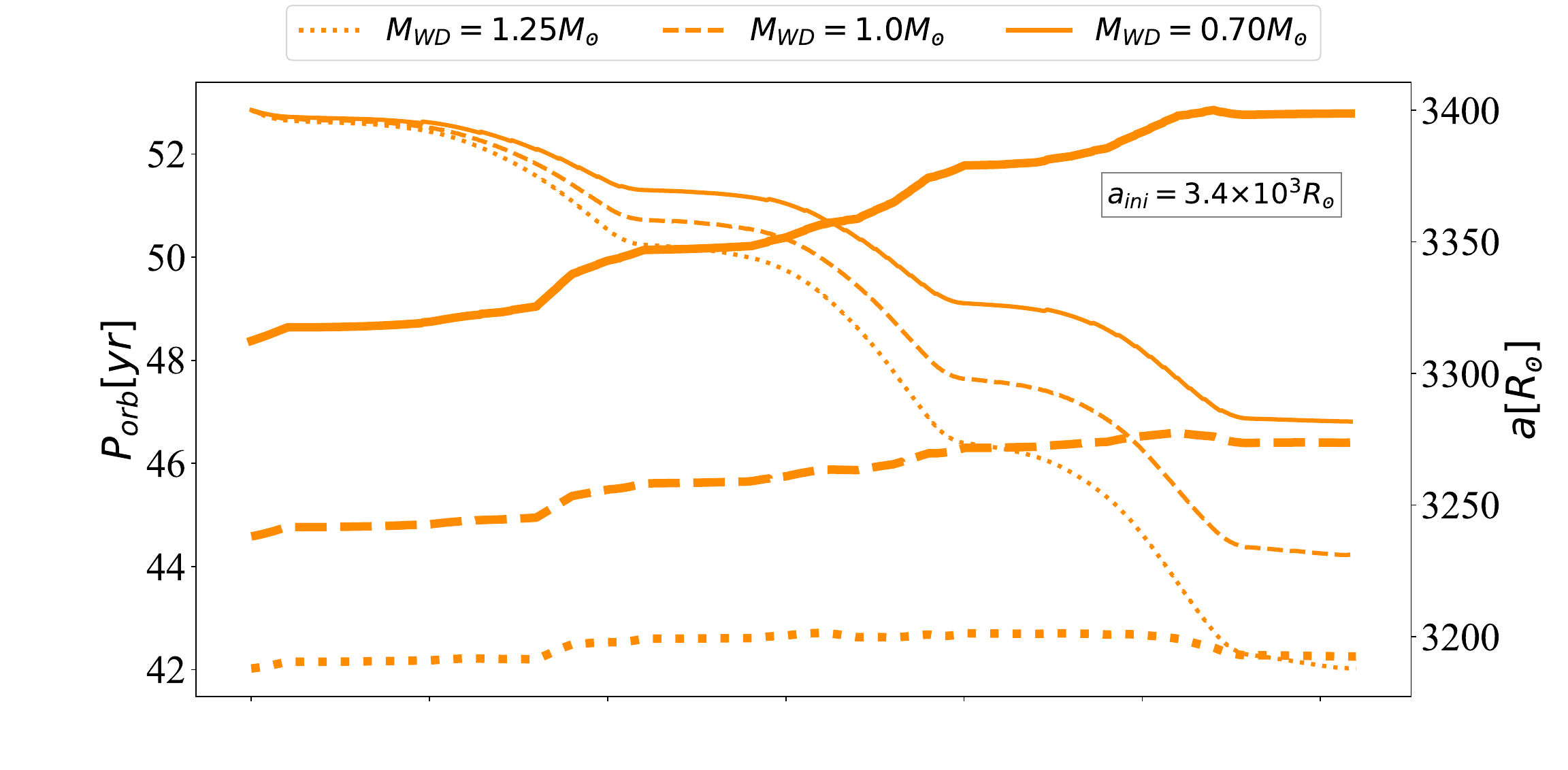}
\end{subfigure}
\hspace*{-0.4cm}
\begin{subfigure}[b]{0.5\textwidth}
   \includegraphics[trim={0.0cm 0.0cm 0.0cm 0.0cm},clip,width=1.04\columnwidth]{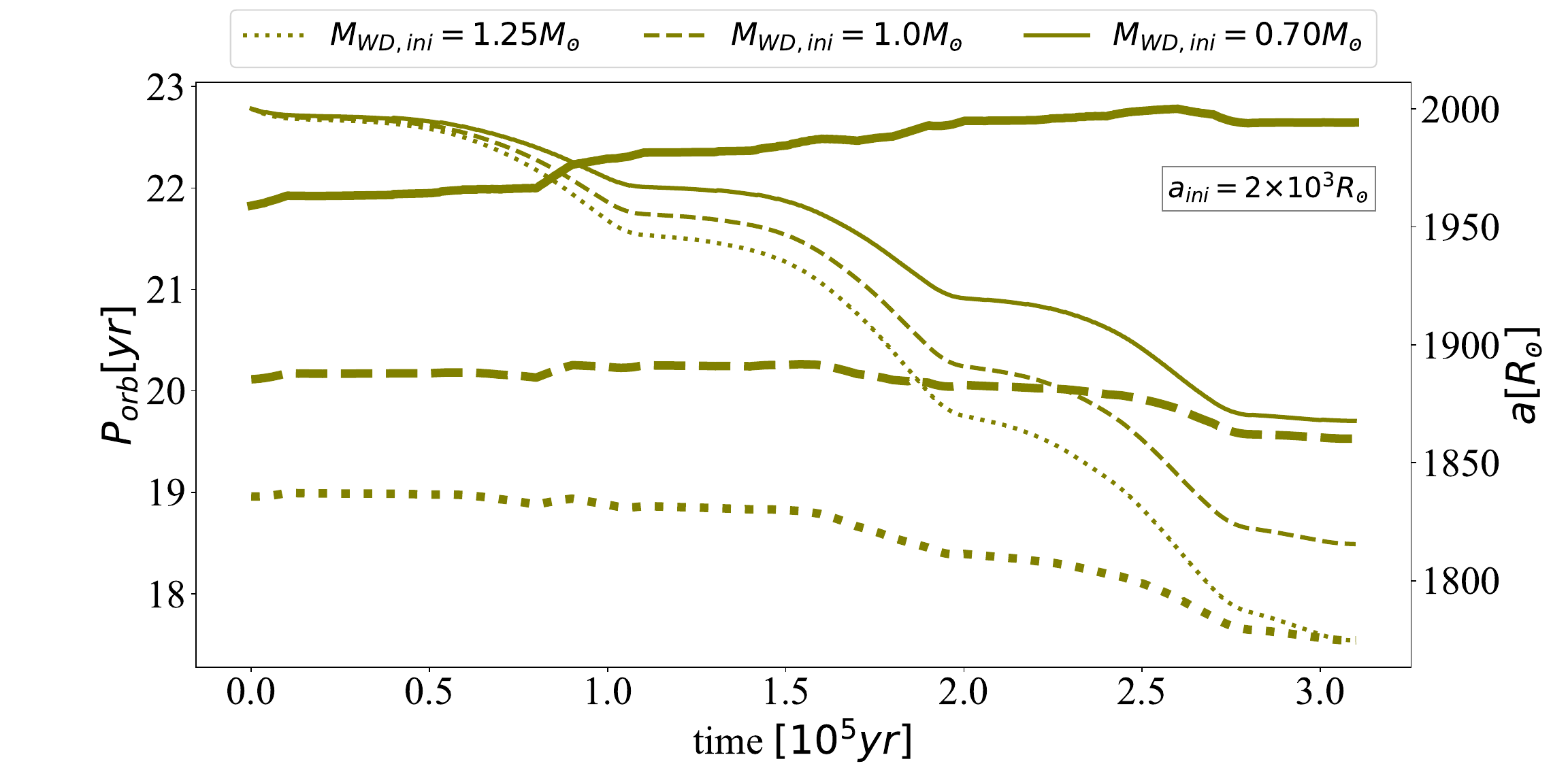}
\end{subfigure}

\caption{Orbital period of various model along with separation plot of the  the binary models are shown in the Figure. The left axis represent the orbital period in years and the right axis represents the change in separation in $\mathrm{R_\odot}$. The color code of WD models are: green-$1.25\mathrm{M_\odot}$; orange-$1.0\mathrm{M_\odot}$ and pink-$0.70\mathrm{M_\odot}$. Line type as described in Figure \ref{fig:Donor and WD}. The first row represents when the separation is $a=2\times10^3 \mathrm{R_\odot}$ (top); second row is $a=3.4\times10^3 \mathrm{R_\odot}$ (middle) and third row is $a=5\times10^3 \mathrm{R_\odot}$ (bottom). The binary separation consistently decreases in all models due to significant AML. Conversely, changes in the orbital period are influenced by variations in both the WD and donor masses, as well as the separation between them. This explains the observed patterns of both increasing and decreasing orbital periods in the Figure. }\label{fig:porb}
\end{figure}

\begin{figure*}
\hspace*{-1.5cm}
   \includegraphics[trim={0.0cm 0cm 0.0cm 0.0cm},clip,width=2.5\columnwidth]{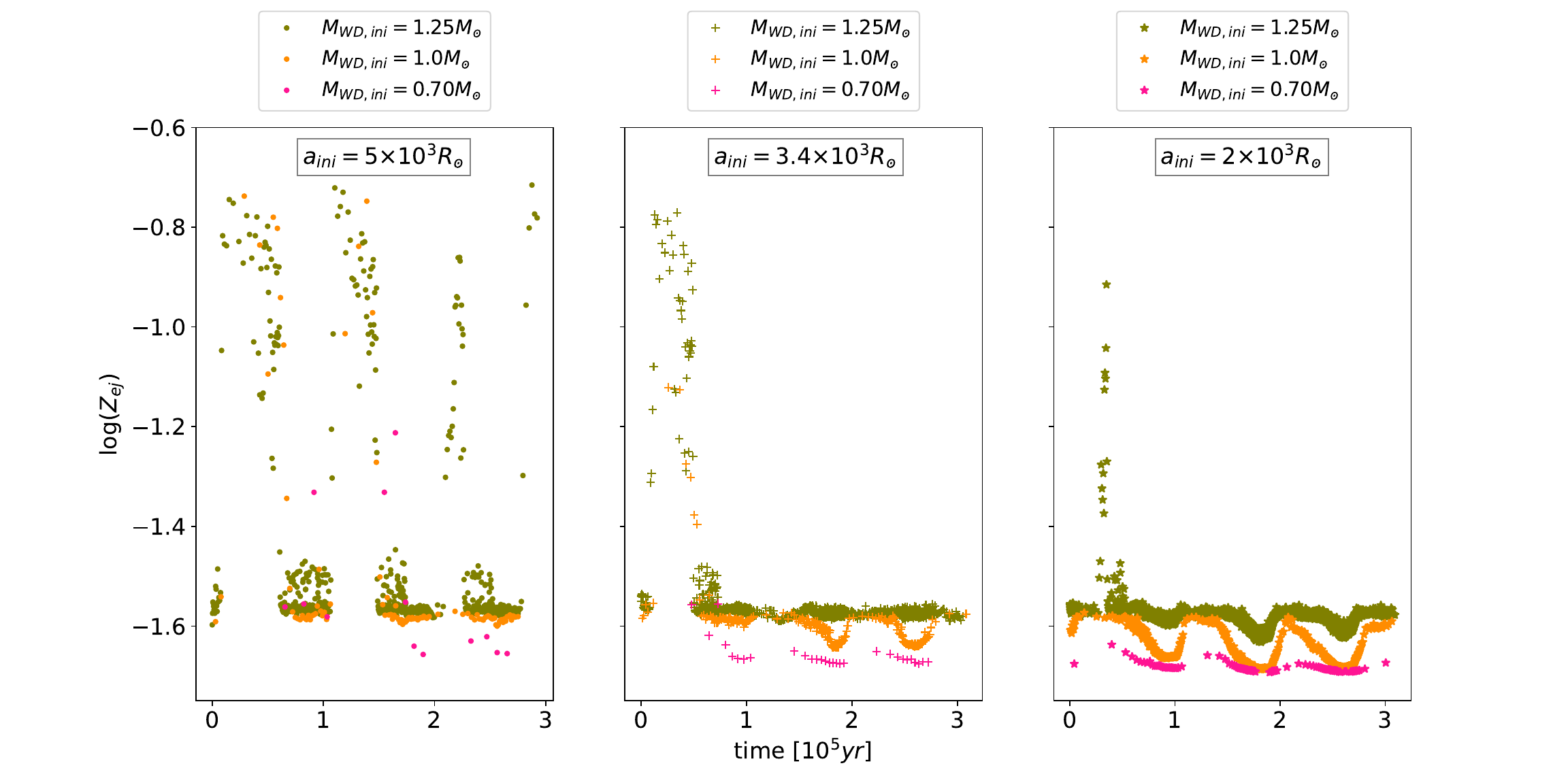}
        \caption{Mass fraction of heavy elements in the ejecta $\mathrm{Z_{ej}}$ on a logarithmic scale vs time. Left: $a=5\times10^3 \mathrm{R_\odot}$; Middle: $a=3.4\times10^3 \mathrm{R_\odot}$; Right: $a=2\times10^3 \mathrm{R_\odot}$. The color code of the WD models are: green-$1.25M_\odot$; orange-$1.0M_\odot$ and pink-$0.70M_\odot$. Abundance is more pronounced in eruptions characterized by lower $\eta$ values, as higher ratios are linked to lower accretion rates. This provides more time for diffusion, which, in turn, leads to the ignition of the TNR occurring at a greater depth beneath the surface.}\label{fig:Z}
\end{figure*}

\begin{figure}
    \hspace*{-0.7cm}
    \includegraphics[trim={0cm 1cm 0.0cm 1cm},clip,width=1.08\columnwidth]{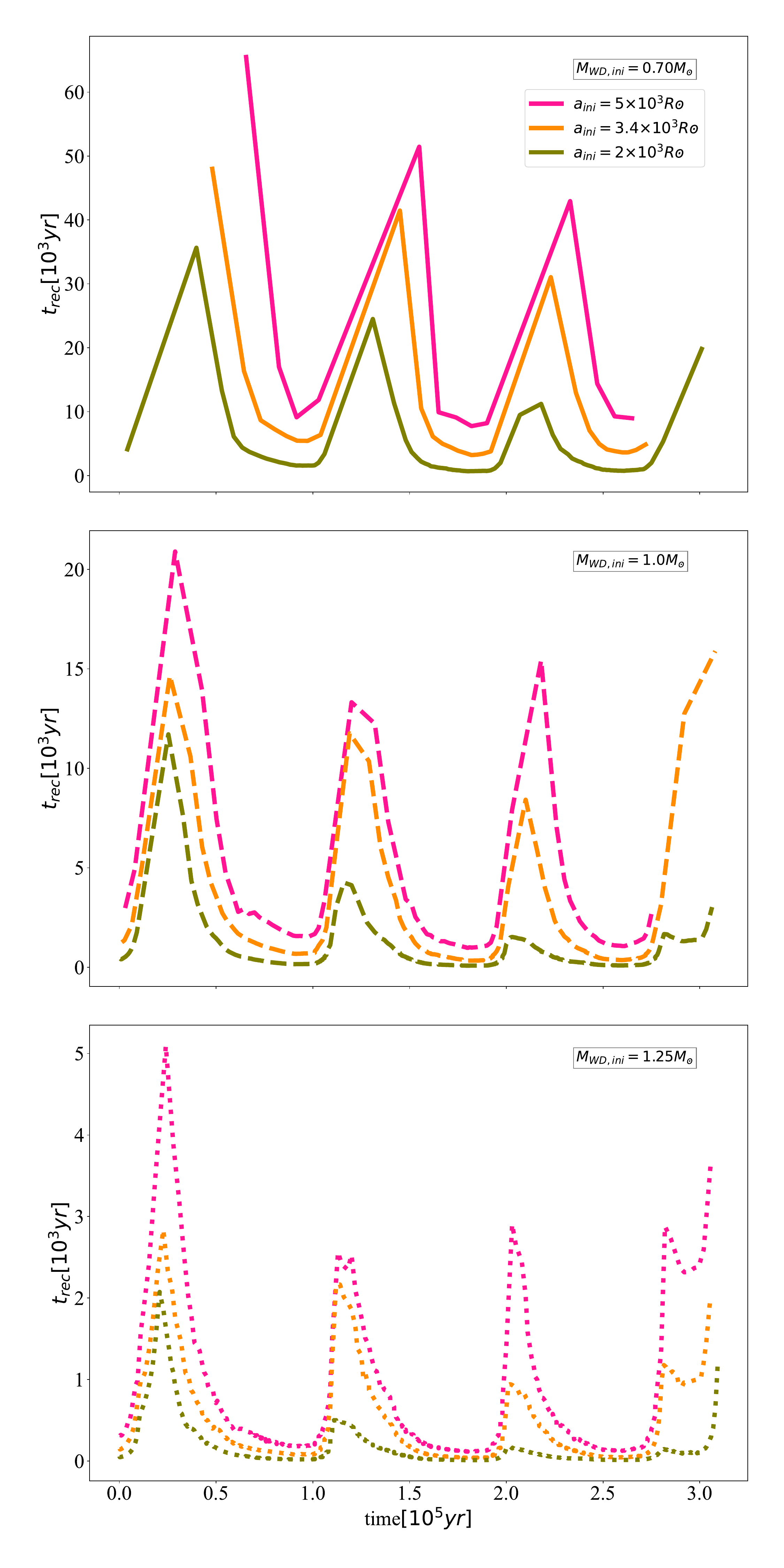}
    \caption{Figure shows the recurrence time for the nine models. Top panel shows the recurrence time of $0.70\mathrm{M_\odot}$ WD model with different separations followed by middle panel showing the same for $1.0\mathrm{M_\odot}$ and bottom panel for $1.25\mathrm{M_\odot}$. The time range of $\mathrm{t_{rec}}$ extends from tens to tens of thousands of years. Across a fixed separation, $\mathrm{t_{rec}}$ consistently exhibits shorter durations for more massive WDs. This occurs because a larger WD requires less mass to attain the triggering mass threshold. The fluctuations in $\mathrm{t_{rec}}$ do not arise due to alterations in separation; instead, they closely follow the pattern of the wind rate trend. The color code and line-type is same as of Figure \ref{fig:Donor and WD} }
    \label{fig:trec1}
\end{figure}

\begin{figure*}
\hspace*{-1.8cm}
   \includegraphics[trim={0.0cm 0cm 0.0cm 0cm},clip,width=2.5\columnwidth]{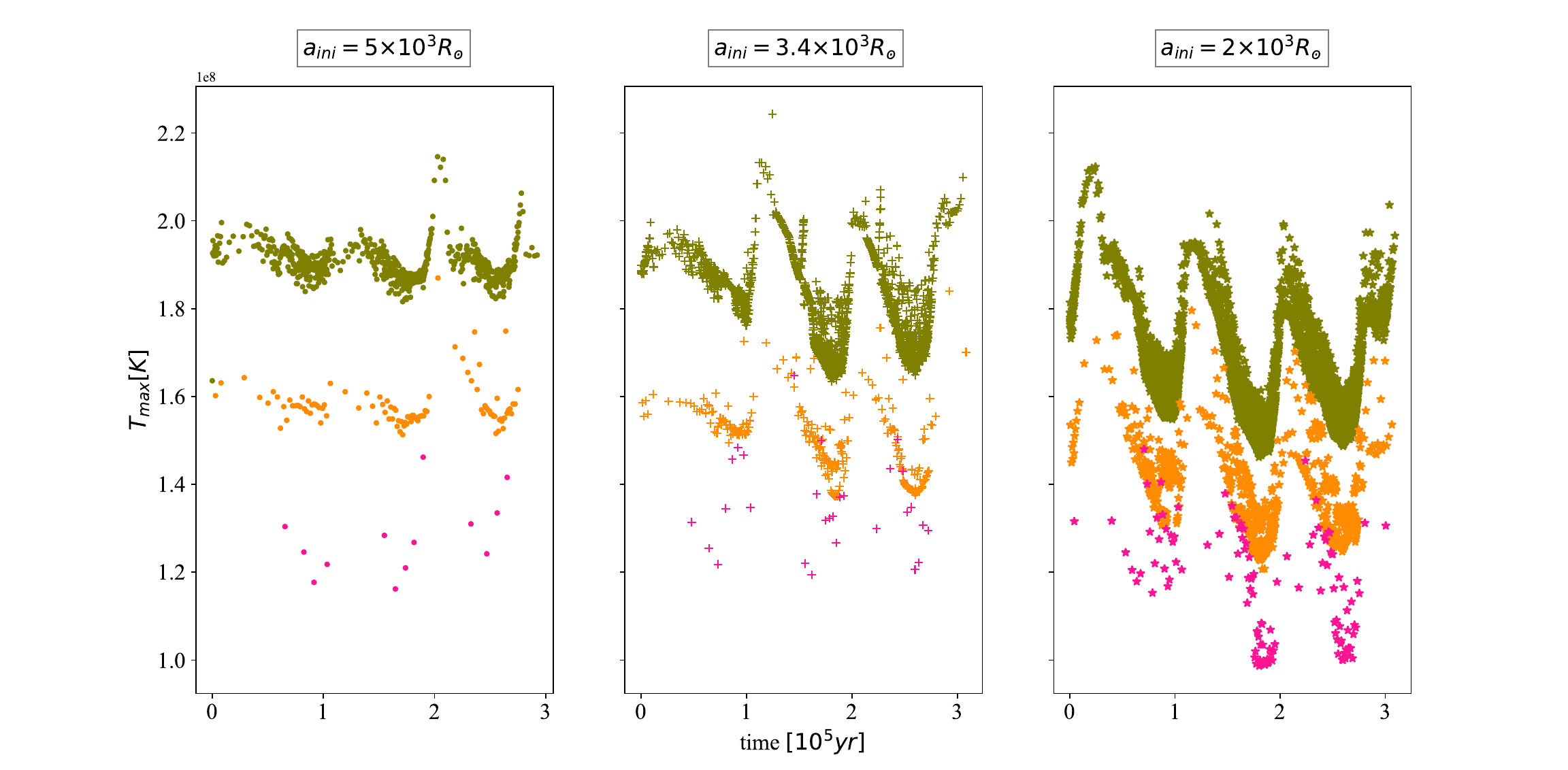}
    \caption{Maximum temperature ($\mathrm{T_{max}}$) attained at each eruption. Left: $a=5\times10^3 \mathrm{R_\odot}$; Middle: $a=3.4\times10^3 \mathrm{R_\odot}$; Right: $a=2\times10^3 \mathrm{R_\odot}$). The color code of WD models are: green-$1.25\mathrm{M_\odot}$; orange-$1.0\mathrm{M_\odot}$ and pink-$0.70\mathrm{M_\odot}$. The $\mathrm{T_{max}}$ shows the maximum value for the massive WDs with the least separation.}\label{fig:tmax}
\end{figure*}

\begin{figure*}
\hspace*{-3.5cm}
\begin{subfigure}[b]{1\textwidth}
   \includegraphics[trim={0.0cm 3cm 0.0cm 1.0cm},clip,width=1.22\columnwidth]{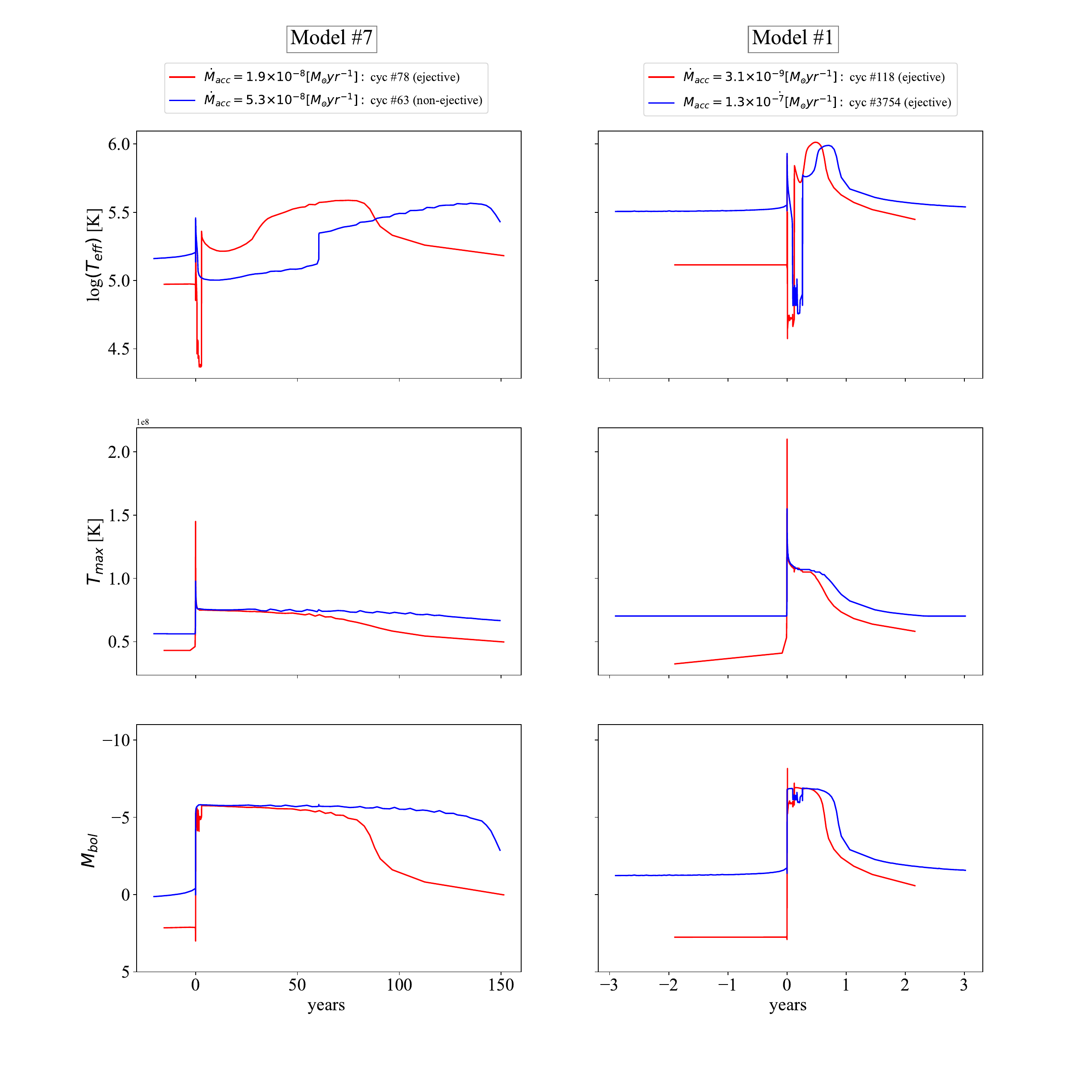}
\end{subfigure}
\caption{Comparison plot of non-ejective (cycle 63) and ejective (cycle 78) WD of model 7 to model 1 having high (cycle 3754) and low (cycle 118) accretion rates the basis of effective temperature (top), maximum temperature (middle) and bolometric magnitude (bottom). The top panel of the figure illustrates that the cycle with a lower $\mathrm{\dot{M}_{acc}}$ cools to a lower $\mathrm{T_{eff}}$ compared to the high $\mathrm{\dot{M}_{acc}}$ cycle, owing to their intense burning and expansion, whereas this contrast is more pronounced in the cycles of model $\#7$. The middle panel of the figure demonstrates the anti-correlation between $\mathrm{T_{max}}$ and $\mathrm{\dot{M}_{acc}}$ — indicating a higher $\mathrm{T_{max}}$ for the lower accretion rate, with both being higher than the $\mathrm{T_{max}}$ observed in model $\#7$. In the lower panel, the amplitude of bolometric magnitude achieved during the eruption is greater for lower $\mathrm{\dot{M}_{acc}}$, attributed to extended diffusion time and more intense TNR. }\label{fig:ejective and non ejective}
\end{figure*}
 
For each of the nine models, we recorded throughout evolution, the number of cycles, the accretion rate ($\dot{M}_{\mathrm{acc}}$), wind rate ($\dot{M}_{w}$), orbital period ($P_{\mathrm{orb}}$), separation ($a$), change in mass of the AGB donor and WD accretor, total accreted and ejected masses ($m_{\mathrm{acc}}$ and $m_{\mathrm{ej}}$ respectively) of each cycle, recurrence time ($t_{\mathrm{rec}}$), maximum temperature ($T_{\mathrm{max}}$) attained during each eruption and the heavy elements abundance ($Z_{\mathrm{ej}}$) in the ejecta. We find many differences between the models, as well as many trends which we specify and elaborate below.  

The evolutionary mass track as well as the radius of the donor AGB from the beginging of the simulations ($t=0$) until their termination is given in Figure \ref{fig:Donor and WD} (top panel). After a time period of about $\sim3\times10^5$ yrs, the donor has finished blowing away its envelope, thus the wind rate has been reduced to a negligible value and the donor mass has been eroded down to  $\sim0.57\mathrm{M_\odot}$. This means a change in the donor mass of $\sim0.43\mathrm{M_\odot}$ i.e., $\sim43\%$ of its initial mass. Examining the evolution of the AGB radius reveals that the donor undergoes periods of expansion and contraction due to thermal pulses, resulting in fluctuations in the wind rate (see Figure \ref{fig:mdot}).
The other three panels of Figure \ref{fig:Donor and WD} show the mass tracks of the WDs for initial masses of 0.7, 1.0 and 1.25$\mathrm{M_\odot}$ exhibiting a similar trend for the three initial WD masses --- a slight decrease in mass, for the largest initial separation i.e., ($a=5.0\times10^3\mathrm{\mathrm{R_\odot}}$), an increase in mass for $a=3.4\times10^3\mathrm{\mathrm{R_\odot}}$, and an increase of about ten times higher for $a=2.0\times10^3\mathrm{\mathrm{R_\odot}}$. The mass differences ($\Delta M_{WD}$) are listed in Table \ref{tab:Table 2}. This dependence of mass growth on separation is the direct result of a smaller separation leading to a higher accretion rate which leads to a more efficient mass retention --- i.e., for a given WD mass, a higher accretion rate leads to less ejected mass during the nova eruption, allowing the WD to have a larger net mass growth at the end of each cycle \cite[]{1994ApJ...424..319K,1995ApJ...445..789P,2005ApJ...623..398Y,2012BaltA..21...76S,2015MNRAS.446.1924H,2021MNRAS.505.3260H}.The total number of cycles per simulation (see Table \ref{tab:Table 2}) supports this trend as well --- more eruptions occur for a smaller separation which means that the average time between two consecutive eruptions decreases with decreasing separation and less time between eruptions means a higher average accretion rate (for a given WD mass). Both of these trends --- of $\Delta M_{WD}$ and of total number of eruptions --- are the direct result of the separation and the accretion rate being anti-correlated. 

The evolutionary behaviour of the accretion rate for the nine models is given in Figure \ref{fig:mdot}. The initial accretion rate ranges from $\sim0.5\times10^{-8}\mathrm{M_\odot yr^{-1}}$ for the least massive WD with the largest separation (model $\#$9), to $\sim6.5\times10^{-8}\mathrm{M_\odot yr^{-1}}$ for  for the most massive WD with the smallest separation (model $\#$1) --- an order of magnitude difference. Examining each WD mass separately for the same separation, and each separation separately for the same WD mass, we see a correlation between the accretion rate and the separation: decreasing the separation or increasing the WD mass both causes the accretion rate to increase. The reasons for these correlations can be understood from Equations \ref{6}  --- which shows that $M_{acc}$ is proportional to $M^2_{WD}$ and inversely proportional to $a^2$. This trend of $\dot{M}-a-M_{WD}$ correlations remains in tact throughout evolution, while all nine curves rise and drop in accordance with the rises and drops of the wind rate. 

Further examining Figure \ref{fig:mdot} reveals that while all the models show the same trend of fluctuations in the accretion rate, the highest momentary rate is obtained for model $\#1$ and the lowest momentary rate is obtained for model $\#9$ --- correlated with the initial values, and continuing this correlation throughout evolution. However, the accretion rate varies constantly between eruptions. Thus to better understand the influence that the accretion rate may have on the outcome of the eruption, we calculate an average accretion rate per cycle. In Table \ref{tab:Table 2} we specify the maximum and minimum average accretion rates per cycle ($ \overline{\dot{M}}_{\mathrm{acc,max}}$ and $ \overline{\dot{M}}_{\mathrm{acc,min}}$ respectively) and the ratio between them, showing that the maximum average accretion rate per cycle is generally between $\sim 25$ and $\sim 75$ times higher than the average minimum depending on the model. Both the average maximum and average minimum span a range of about one order of magnitude between the different models, with the highest average maximum rate being for model $\#1$ (the most massive WD and the smallest separation) and the lowest average minimum rate being for model $\#9$ (the least massive WD and the largest separation). Since the WD mass and binary separation have competing effects on the accretion rate (as explained earlier), we find a coincidental, but remarkable overlap between the accretion rates of the $1.0\mathrm{M_\odot}$ model with the $5.0\times10^3\mathrm{\mathrm{R_\odot}}$ separation and the $0.70\mathrm{M_\odot}$ model with the $3.4\times10^3\mathrm{\mathrm{R_\odot}}$ separation (models $\#6$ and $\#8$). Additionally, the accretion rate evolution shows that for wider separations the fluctuations are substantially less pronounced than they are for smaller separations.

We define the mass \textit{transfer} efficiency as the ratio between the accretion rate and the wind rate ($\zeta=\dot{M}_{acc}/\dot{M}_w$) and show this ratio vs. time in Figure \ref{fig:Accretion rate}, demonstrating how much mass is actually captured by the WD via the BHL mechanism. The value of $\zeta$ increases with increasing WD mass (shown by the orange line being higher than the pink, and the green being higher than both the orange and the pink, in all three panels of Figure \ref{fig:Accretion rate}). Additionally, $\zeta$ also decreases with increase in separation (demonstrated by the same color line getting higher when moving from the left panel to the right). The highest accretion efficiency is $\sim5\%$ and the lowest is only $\sim0.2\%$, meaning that at least $95\%$ of the mass blown away from the AGB is not captured by the WD. Our accretion efficiencies are in agreement with the range obtained by \cite{2009ApJ...700.1148D} for widely separated systems. 

Figure \ref{fig:mejtomacc} shows the accreted mass ($m_{\mathrm{acc}}$) per cycle; the ejected mass ($m_{\mathrm{ej}}$) per eruption; and the mass \textit{retention} efficiency, which we define as: $\eta=(m_{acc}-m_{ej})/m_{acc}$ giving the ratio of the net amount of accreted mass retained on the WD after eruption to the total amount of accreted mass per cycle. The accreted masses are higher for less massive WDs, and the ejected mass is greater for wider separations. This is because of the anti-correlation between the separation and the accretion rates --- both trends which are consistent with other authors  \cite[]{1995ApJ...445..789P,2005ApJ...623..398Y}. For reference, we specify in Table \ref{tab:Table 2} the maximum and minimum accreted mass per cycle for each model. The maximum accreted mass being correlated with the minimum average accretion rate and minimum accreted mass being correlated with the maximum average accretion rate. We also find that the mass \textit{retention} efficiency, $\eta$, decreases with increasing separation, its maximum possible value being equal to 1 for rare non-ejective cycles. Negative values of $\eta$ are in correlation with lower accretion rates.  This is in agreement with previous works \cite[]{2005ApJ...623..398Y,2014ASPC..490..287N,2021MNRAS.505.3260H}.

In \cite{2016ApJ...819..168H} it is shown that there is a limit of a few times $10^{-8}\mathrm{M_\odot yr^{-1}}$ under which $m_{\mathrm{ej}}>m_{\mathrm{acc}}$ and above which $m_{\mathrm{ej}}<m_{\mathrm{acc}}$. For our most widely separated models, the average accretion rates are mostly below this limit for the entire evolution (except for a few cycles, as expressed in $\eta$ being below 0), leading to secular mass loss, whereas for the other separations, the average accretion rates are above this limit for part of the evolution and below for part, leading to $\eta$ alternating below and above 0 depending on the average accretion rate for the specific cycle, but ultimately leading to a net mass gain. This may be seen in $\Delta M_{WD}$ in Table \ref{tab:Table 2}.  

From Figure \ref{fig:mejtomacc} it is obvious that the mass retention efficiency ($\eta$) changes from one eruption to the next --- in stark contrast with the slow, secular, unnoticeable changes typical of CVs in RLOF \cite []{2020NatAs...4..886H,2021MNRAS.505.3260H}. This is due to the change in the average accretion rate from one cycle to the next, which is caused by the changing wind rate. For smaller separations, the $\eta$ curves typically show higher values, mostly above 0, which aligns with the positive $\Delta \mathrm{M_{WD}}$ values in Table \ref{tab:Table 2}, while showing the opposite trend for larger separations. For our least massive WD with the smallest separation (model $\#7$) this ratio is 1 for some cycles, meaning that there was no mass ejection during these novae eruptions. The average accretion rate that yeilded non-ejective cycles is of the order $\approx5\times10^{-8}\mathrm{M_\odot yr^{-1}}$.

This is a case where a TNR that occurs on the surface of a WD leads to the ejection of very little mass or none at all. There is a narrow range of accretion rates for which such mild TNR can occur without mass ejection as given by \cite[]{2005ApJ...623..398Y} which are cases with considerably high accretion rates. In such cases the mass is accreted so rapidly, that there is very little time for diffusion, thus the TNR occurs close to the surface, entailing less fusion thus initiating a weak TNR that is not capable of producing sufficient energy to bring the ejecta to the escape velocity. Thus, the outer layers of the WD expand, causing a brightening (the nova), the burning is then quenched, and the envelope contracts because the energy is consumed \cite[]{2019ApJ...879L...5H}. \cite{2013ApJ...777..136W} obtained periodic hydrogen flashes for accretion rates below the stable burning threshold (a state in which the rate that hydrogen burns is exactly equal to the rate that it is accreted). When the rates were slightly below to the stable burning threshold, they obtained flashes with minimal mass loss from the system. 
In our code, the high accretion rates that give us non-ejective outbursts are lower for less massive WD’s and overall similar to the steady-burning rates of \cite{2007ApJ...660.1444S} and \cite{2013ApJ...777..136W}. The results are fairly similar - they obtain continuous burning with no eruptions, while we obtain a regime where the eruption begins, and then the hydrogen burns before becoming energetic enough to eject mass, i.e., the hydrogen is not burnt at a constant rate, but in short weak epochs. Differences between the results of the two codes may be a consequence of different architectures and assumptions. Since the WD brightens and begins to expand (and briefly stops accreting), we define it as a weak nova eruption in which no mass is ejected. Both codes support this regime of high accretion rates to be the regime within which we should be searching for SNIa progenitor candidates because the ultimate outcome is an increase in WD mass. \cite{2019ApJ...879L...5H} have obtained non-ejective eruptions for a WD mass of $1.1\mathrm{M_\odot}$, when the accretion rate  $\approx 3.5-5 \times10^{-7}\mathrm{M_\odot yr^{-1}}$. Here we see that for our higher WD masses, the accretion rate becomes higher than for our $0.70\mathrm{M_\odot}$ WD, yet the former did not become non-ejective at any point, and the latter did. This indicates that the lower limit on the average cyclic accretion rate, for producing non-ejective novae is lower for lower WD masses.

The change in the separation as well as the change in the orbital period throughout the evolution are given in Figure \ref{fig:porb}. All the models show a similar trend of consistent decrease in the separation with time as seen in long-term modelling of CVs. We find this trend to be steeper for WDs with the smallest separation, because such models experience the most AML. For a given separation more massive WDs lose more angular momentum than less massive ones because the change in angular momentum due to magnetic braking ($\dot{J}_{MB}$) is dependant on $P_{\mathrm{orb}}^{-3}$ --- i.e., proportional to $M_{WD}^{3/2}$ (Equation \ref{7}). Similarly, angular momentum due to gravitational braking ($\dot{J}_{GR}$) is dependant on $P_{\mathrm{orb}}^{-7/3}(M_{WD}^2/M_{WD}^{2/3})$ --- i.e., proportional to $M_{WD}^{5/2}$ (Equation \ref{8}) and the momentum change due to drag is dependant on $r_a^2$ --- i.e., proportional to $M_{WD}^{2}$ (Equation \ref{9}). All these equations indicates that the mechanism of AML is stronger for massive WDs thus occurs at a higher rate. 
 
We see, in stark contrast with the orbital period evolution of novae in CVs, that the orbital period does not always follow the trend of the separation--- and there are cases for which the trend is reverse. This can be explained according to Kepler's third law $P_{\mathrm{orb}}^2$=\(4\pi G a^{3}/(M_{WD}+M_{AGB})\), the orbital period depends on the mass of the system as well as their separation. For novae in CVs, the mass of the system changes very slowly, so the decreasing separation leads to a decreasing orbital period. However, here the companion is an AGB donor that loses mass rapidly, which affects the orbital period by increasing it. This yields a situation in which both the separation and the total mass of the system decrease over time, but they have competing influences on $P_{\mathrm{orb}}$ -- decreasing the separation decreases $P_{\mathrm{orb}}$ whereas decreasing the total mass increases $P_{\mathrm{orb}}$. This means that the period can increase even if the separation decreases, if the change in mass is faster than the change in separation i.e., $\Delta a^3/a^3 > \Delta M_{AGB}/M_{AGB}$ it will lead to a decrease in $P_{\mathrm{orb}}$, whereas if $\Delta a^3/a^3 < \Delta M_{AGB}/M_{AGB}$ it will lead to an increase in $P_{\mathrm{orb}}$. We find that for the widest separation, the orbital period increases for all three WD mass models, throughout the evolution. This is because the relative separation changes for this initial separation is the smallest, hence the mass change is dominant, leading to an increase in the orbital period. For the two other initial separations we find that some lead to an increasing $P_{\mathrm{orb}}$ and some leads to a decreasing $P_{\mathrm{orb}}$, depending on the model (see Figure \ref{fig:porb}).
 
We turn to investigate the composition of the ejected material (Figure \ref{fig:Z}), and find that the abundance of heavy elements in the ejecta ranges from $\sim 2\%$ to $\sim 20\%$. This is anti-correlated with the ratio of the net-accreted to ejected mass per cycle, $\eta$ (Figure \ref{fig:mejtomacc}), a higher abundance is present in eruptions with a lower $\eta$ --- in agreement with \cite{2022MNRAS.511.5570H}. This happens because high ratios occur for low accretion rates, thus there is more time for diffusion and mixing of the accreted envelope with deeper, more enriched (in CNO mostly) layers, the TNR is ignited deeper below the surface, thus, the resulting ejecta contains a higher abundance of heavy elements \cite[]{1984ApJ...281..367P,1992ApJ...388..521I,1995ApJ...445..789P,2005ApJ...623..398Y}.

An obvious anti-correlation that we find is of the recurrence period (Figure \ref{fig:trec1}) with the accretion and wind rates (Figure \ref{fig:mdot}). This is simply since, for a given WD mass, a higher accretion rate allows reaching the required critical amount of mass that is needed for initiating a TNR in less time. The recurrence period (Figure \ref{fig:trec1}) also follows the trend we see in the ejecta abundance, where shorter (longer) recurrence periods lead to lower (higher) enrichments, this is because a longer time between eruptions allows more time for mixing as explained above. The range of \(t_{\mathrm{rec}}\) is from tens to ten thousands of years. For a given separation, $t_{\mathrm{rec}}$ is consistently shorter for more massive WDs. This is because, a more massive WD needs less mass to acquire the triggering mass, and also it has a larger accretion radius, and thus a higher accretion rate than less massive WDs for the same wind rate. We note that the fluctuations in the $t_{rec}$ are not caused by the change in separation but rather it follows the trend of the wind rate. 

The maximum temperature ($T_{\mathrm{max}}$) of the nine models ranges from $1$--$2.2\times10^{8}$K, as depicted in Figure \ref{fig:tmax}, following the same trend as the $Z_{\mathrm{ej}}$ --- higher temperature for more massive WDs and for a bigger separation, although the correlation with separation is less pronounced because although it effects the accretion rate, the accretion rate has a secondary influence on the $T_{\mathrm{max}}$ while the key parameter is the WD mass. The fluctuations are anti-correlated with the accretion rate --- a higher temperature for lower accretion rates in agreement with models of novae in CVs, (e.g., \cite{1995ApJ...445..789P,2005ApJ...623..398Y}).

We return to discuss the non-ejective cycles of model $\#7$. As explained earlier, they are the result of a relatively high accretion rate. In order to demonstrate the observational difference between such an eruption and a "typical" eruption, we compare one of these non-ejective cycles (cycle $\#63$) with a different ejective cycle (cycle $\#78$) from a later point in the evolution of the same model. The average accretion rate of cycle $\# 63$ is $\approx 5 \times 10^{-8}\mathrm{M_\odot yr^{-1}}$ and of cycle $\# 78$ is $\approx 1 \times 10^{-8}\mathrm{M_\odot yr^{-1}}$. We show the comparison of the effective temperature $(T_{\mathrm{eff}})$, maximum temperature ($T_{\mathrm{max}}$) and bolometric magnitude ($M_{\mathrm{bol}}$) of the two cycles in Figure \ref{fig:ejective and non ejective}.
We find that for the non-ejective eruption the effective temperature of the WD is higher than that of the cycle in which there is mass ejection --- $\sim 1\times10^5$ K as apposed to $\sim 2.5\times10^4$ K. This is because of the nature of the evolution of the effective temperature during a nova cycle. While the WD is quiescently accreting mass, the effective temperature is high --- the surface temperature of a WD. During an eruption, the outer layers expand and cool, lowering the effective temperature. If the eruption is weak, then the outer layers do not expand as much as for a strong ejection. For a non-ejective eruption there is very little expansion, thus the effective temperatute remains high. For the same reason the $T_{\mathrm{max}}$ is higher for the ejective cycle --- because there is substantial burning, whereas in the non-ejective cycle there was very little burning. We note that during quiescence, both of these temperatures ($T_{\mathrm{eff}}$ and $T_{\mathrm{max}}$) are higher for the non-ejective cycle. This is because the time between eruptions ($t_{\mathrm{rec}}$) is shorter thus the WD has less time to cool before erupting again. The bottom panel of Figure \ref{fig:ejective and non ejective} shows the bolometric magnitude for the two cycles, showing the maximum value --- during eruption and decline --- to be the same for both cycles --- corresponding to a few times the solar luminosity which is the Eddington limit. Since we assume black body radiation, and the effective temperature during eruption is different for these two cycles, we may deduce that the eruption for which the WD's $T_{\mathrm{eff}}$ is lower --- the ejective cycle --- will be brighter in the visual than the non-ejective cycle which should have a brighter magnitude in the UV. We also stress that the amplitude of the eruption for the non-ejective cycle is smaller because it was higher in quiescence (during accretion phase). This is a result of accretion taking place at a rapid pace, thus the WD did not have much time to cool down from the previous cycle as it would in cycles with lower accretion rates, such as cycle $\#78$ which has an amplitude of $\sim \times2.5$ the amplitude of cycle $\#63$.

The three right panels of Figure \ref{fig:ejective and non ejective} show such a comparison for two cycles from the simulation with the same separation, but a WD mass of $1.25\mathrm{M_\odot}$( model $\#1$). We compare between the two cycles  with the highest (cycle $\#118$) and the lowest (cycle $\#3754$) average accretion rates. During the eruption, the cycle with the lower accretion rate reaches a lower $T_{\mathrm{eff}}$ than the cycle with the high accretion rate. This is, again, a direct cause of the cooling time between eruptions, although since both of these cycles experience intense burning and expansion, the difference between them is not as stark as for the cycles of model $\#7$. Additionally, the effective temperature of model $\#1$ is higher than that of model $\#7$ right after the eruption and as it declines because of the more massive WD allowing less time for cooling between eruptions. The anti-correlation between $T_{\mathrm{max}}$ and $\dot M_{\mathrm{acc}}$ is well observed in Figure \ref{fig:ejective and non ejective}, as well --- a higher value of $T_{\mathrm{max}}$ for the lower accretion rate, and both are higher than the $T_{\mathrm{max}}$ of model $\#7$. The maximum bolometric magnitude is around $-8.4$ ($-6.8$) for low (high) accretion rates, which brighter than the cycles from model $\#7$ (-5.7). Comparing between the two cycles of model $\#1$, we find that the eruption amplitude is 5.4  for the high accretion rate and 11.1 for the low accretion rate --- twice the amplitude. This is attributed, as with model $\#7$ to the longer time between eruptions that allows for cooling and the model being able to retain some of the accreted mass, and growing in size. 

\section{Discussion }\label{sec:discussion}

For novae that occur in CVs with red dwarf (RD) donors, the mass transfer is powered by RLOF, causing the accretion rate to respond exponentially to minuscule separation changes \cite[]{1988A&A...202...93R} as well as being enhanced temporarily after each eruption due to irradiation of the donor due to eruption \cite[]{1988ApJ...325..828K,1999MNRAS.310..398I,2001A&A...375..937S,2008A&A...483..547T,2011ApJS..194...28K,2014ApJ...784L..33G,2020NatAs...4..886H,2021MNRAS.507..475G, 2021MNRAS.505.3260H}. We do not see this behavior in widely separated symbiotic systems because the wide separation avoids the substantial irradiation of the donor by the WD during eruption, and more importantly, because the mass transfer is not powered by RLOF, avoiding the exponential dependence. Still, we find that the separation, the mass of the WD and the accretion rate are all correlated. The accretion rate decreases with the decrease in the mass of the WD and increases with the decrease in the binary separation. Nevertheless, even though the mechanism that determines the mass transfer rate in these widely separated symbiotic systems is entirely different than the mechanism in CVs and it responds differently to the changing separation, the outcome of the nova - namely, the critical mass required to initiate the TNR, the amount of ejected mass, the magnitudes, temperatures and recurrence times, are basically the same for both types of systems. 

One of our results that was not attained for the group of self-consistent RLOF CV models with zero age main sequence (ZAMS) donors in \cite{2021MNRAS.505.3260H} are non-ejective cycles. Their parameter combinations did not permit the accretion rate to become high enough. Here, for our smallest separation with a WD of $0.7\mathrm{M_\odot}$, we obtain at portions of the evolution, accretion rates high enough to produce non-ejective cycles, implying that systems with AGB donors may be capable of producing higher accretion rates than systems with RD donors. Another interesting point here is that these non-ejective cycles occurred for an accretion rate of $\sim 5\times 10^{-8} \mathrm{M_\odot yr^{-1}}$, whereas more massive WDs attained a higher rate ($\sim 1\times 10^{-7} \mathrm{M_\odot yr^{-1}}$), but did not become non-ejective. Non-ejective outbursts have been obtained by \citet{2019ApJ...879L...5H} in a WD with a mass of $1.1\mathrm{M_\odot}$, occurring when the accretion rate is approximately $3.5-5 \times10^{-7}\mathrm{M_\odot yr^{-1}}$. Similar outbursts were also achieved by \citet{2005ApJ...623..398Y} for WDs with a mass of $0.65\mathrm{M_\odot}$ and an accretion rate ranging from $\times 10^{-6}$ to $\times 10^{-7}\mathrm{M_\odot yr^{-1}}$. According to the analysis, it is theoretically feasible for nova eruptions to occur without significant mass ejection (or none at all) given that the rate of mass accretion is adequately high \cite[]{1995ApJ...445..789P,2005ApJ...623..398Y,2012BASI...40..419S,2019ApJ...879L...5H}.
While examining various models, we find that even though more massive models than $0.70\mathrm{M_\odot}$ attain a higher accretion rate, such systems cannot produce non-ejective eruptions. This helps us to learn that the threshold for the average cyclic accretion rate, below which non-ejective novae are produced, decreases for less WDs, not only in CVs, but also in symbiotic systems. Thus, we expect the same behaviour to reappear for even less massive WD models. Also, these non-ejective cases are growing in mass indicating the possibility of their WDs reaching the Chandrasekhar mass. However, the journey to this ultimate mass is long and many conditions may change. For instance, the average accretion rate might fall below the threshold required for net mass gain, the WD could experience helium flashes leading to a temporary growth cessation until it re-establishes equilibrium \cite[]{2016ApJ...819..168H} or the obvious issue --- the donor simply may not have enough mass to supply. In our low-mass WD with the $1 \mathrm{M_\odot}$ AGB donor, it is likely that the mass source will be exhausted before the WD attains the Chandrasekhar mass.

 If we would want to learn something about the progenitor of a nova based on the observed composition of the ejecta, we would have to account for it being in a symbiotic system. And sometimes this ejecta becomes a favourable candidate for dust formation which may contain silicates, amorphous carbon, silicon carbide and hydrocarbons. Novae V2361 Cyg and V2362 Cyg have such emission spectra due to their carbonaceous grains and the reason for such features is due to the presence of hydrogenated amorphous carbon \cite[]{2011EAS....46..407H}. The carbon and oxygen formed in the ejecta remain available for dust formation, as the incomplete carbon monoxide does not become saturated \cite[]{1997MNRAS.292..192E}. Nova ejecta's free expansion makes the formation of dust exceedingly difficult, while a radiative shock can make dust production highly efficient \cite[]{2017MNRAS.469.1314D,2019PASP..131l4202D}. This means that the ejecta can be contaminated by its surroundings, and therefore it may be difficult (or impossible) to deduce what originated from the WD and what was collected from interaction with the surroundings, thus, an accurate ejecta composition may remain unattainable, as well as an accurate deduction of the accretion rate.

To estimate the amount of mass needed to grow a WD to the Chandrasekhar limit ($\mathrm{M_{Ch}}$), we calculate the mass fraction of the wind that remains on the WD by taking the product of the mass transfer efficiency, $\eta$, and the mass retention efficiency, $\zeta$. 
The product value decreases with an increase in mass for the separation of ${2\times10^{3}\mathrm{R_\odot}}$ and ${3.4\times10^{3}\mathrm{R_\odot}}$. However, this trend shows an opposite pattern for the models with the least separation. The more massive model ($\#3$) has the maximum value compared with the other two models, ($\#6$ and $\#9$), and the value decreases with a decrease in mass. Therefore, we can conclude that when the product of $\eta$ and $\zeta$ is positive, it results in a net gain in mass or growth of the WD, and a higher value indicates a more efficient outcome. However, it's essential to remember that even when considering the best value of the product of $\zeta$ and $\eta$, which is 0.02 for a $0.70\mathrm{M_\odot}$ WD and $1.0\mathrm{M_\odot}$ AGB, the WD can only grow up to $0.72\mathrm{M_\odot}$.
This represents the maximum mass the WD in this model can attain, assuming the accretion rate remains at this high rate --- which is unrealistic, but even so, the AGB does not have even nearly enough mass to supply to the WD to allow it to become a type Ia supernovae (SNIa). Repeating the same calculations for our model with the best product of $\zeta$ and $\eta$ for our $1.25\mathrm{M_\odot}$ WD, we obtain  the best product value as 0.02 leading to a WD mass of $1.27\mathrm{M_\odot}$ which again shows the inability of these parameters to lead the WD to a SNIa. We stress that the absence of a candidate model for SNIa in this work does not rule out the possibility of it. This result is extremely important because it steers us in the right direction by deducing that a relevant candidate must have more massive stellar components. Let us take as an exercise, a better efficiency, say, 0.05, with a $1.3\mathrm{M_\odot}$ WD and a $2.0\mathrm{M_\odot}$ donor AGB. This will result in the WD growing to $1.4\mathrm{M_\odot}$, i.e., reaching the $\mathrm{M_{Ch}}$. This hypothetical calculation suggests that in order to attain $\mathrm{M_{Ch}}$, an initial massive WD is required. However we stress that a self-consistent model of this is needed in order to asses if these parameters (mostly the efficiency) is naturally attainable.

There is another possibility of recurrent novae eruptions in some of the models. Since the accretion rate increases with a decrease in separation and an increase in the mass of the WD, it favors faster accretion of matter on the surface of the WD, resulting in frequent outbursts and leading to a shorter recurrence time. Here, the shortest recurrence time occurs for the most massive WD with the smallest separation $(\approx 10$ yrs), while the longest recurrence time is observed for the least massive WD with the largest separation $(\approx 6\times10^4$ yrs).
While some of the nova eruptions from the model with the most massive WD and smallest separation (model $\#1$) can be taken into account as eruptions falling into the category of recurrent novae (RN) (with a $t_{\mathrm{rec}}$ < 100 yrs) whereas, the models with the widest separation falls into the category of classical novae (CN).

We have also seen that a WD with a mass of $1.25\mathrm{M_\odot}$ with the smallest separation (model $\#1$) has a better chance of growing in mass by accreting matter, i.e., it is able to accumulate more mass and \textit{might} reach the limiting value of Chandrashekar limit. Any of the models with a positive $\Delta \mathrm{M_{WD}}$ might be candidates for a SNIa progenitor, provided their donor has enough mass to supply. Thus, systems that host massive WDs with giant donors that have enough mass to donate and a relatively small separation, might favour the possibility of SNIa explosion.

There are a few known symbiotic systems that have frequent eruptions, and therefore it is possible that their WD's mass is increasing. For instance, the symbiotic recurrent novae V1016 Cyg is a widely separated binary system, which had a pre-brightening outburst in 1949, followed by a nova eruption in 1964 ($\sim$5-7 mag in the V-band). It was followed by two small outbursts (less bright) in 1980 and 1994, suggesting the system has a recurrence period of $15.1\pm0.2$ years. The system consists of a cool Mira component with a mass of $0.81\pm0.20 \mathrm{M_\odot}$ and a WD of mass $1.1\mathrm{M_\odot}$ \cite[]{1992MNRAS.256..177M,2001OAP....14...61P,2003CoSka..33...99P,ArkhipovaTaranovaIkonnikovaEsipovKomissarovaShenavrin+2016+35+41}.

The RN RS Ophiuchi (RS Oph) consisting of a WD ($\sim 1.2-1.4 \mathrm{M_\odot}$) and a red giant ($0.68-0.80 \mathrm{M_\odot}$) is another example of a RN which had several outbursts. Namely, in 1898, 1933, 1958, 1967, 1985, 2006 and recently in 2021 \cite[]{2008NewAR..52..386H,2009A&A...497..815B,2022BlgAJ..37...24Z} with a shorter recurrence period of 21 years. The lack of any indication of an oxygen-neon (ONe) WD in the UV spectroscopic characteristics suggests that RS Oph  hosts a massive carbon-oxygen (CO) WD. Since a WD with a mass greater than $\sim 1-1.1\mathrm{M_\odot}$ is expected by stellar evolution theory to be ONe, the spectroscopy showing it to be CO indicates that the WD has acquired this mass through its evolution by accreting material from the donor star \cite[]{1996ApJ...460..489R,2017ApJ...847...99M}. Theoretical models suggest that systems like RS Oph could become a SNIa within $10^5-10^7$ years \cite[]{2008NewAR..52..386H}, This is validated by the eruption in 2006, where the mass of the ejected shell was only $\sim10^{-7}\mathrm{M_\odot}$, indicating that the WD grows in its mass \cite[]{2006Natur.442..276S}, similar to our models. However, this ultimately depends on the mass transfer rate remaining as it is, and this is not guaranteed for several reasons, the most important of them being the possibility of the donor running out of mass before the WD reaches $\mathrm{M_{Ch}}$.

Another recurrent novae is V407 Cygni, which had an outburst in 1936-1939. It consists of a WD ($1.2 \mathrm{M_\odot}$) and a Mira-type giant ($1.0 \mathrm{M_\odot}$) with an orbital period of 43 years. It was followed by another small outburst in 1988 and a bright one in 2010 \cite[]{1949vvff.book.....H,1990MNRAS.242..653M,2012MNRAS.419.2329O,2015BaltA..24..353E}.

Eruptions in certain symbiotic systems, specifically those with less massive WD components, can span several decades. Such eruptions fall into the category of CN. Due to their infrequent nature, predicting the rate at which the components gain or lose mass becomes challenging. For instance, AG peg is an example of such a symbiotic system with a red giant ($2.5\mathrm{M_\odot}$) and a WD ($0.6 \mathrm{M_\odot}$) with an orbital period of 818 days, which had an increase in its brightness in mid $19^{th}$ century from $\sim9$ to $\sim6$ mag. Subsequently, it entered a quiescent phase that persisted for roughly 130 years, until another outburst took place in 2015. The former symbiotic nova outburst, which had a profound impact on AG Peg for over a century, came to an end before the occurrence of a distinct outburst in 2015. It is worth noting that the former type of outburst, depending on the mass of the white dwarf and the accretion rate, would endure for many years, leading to a slow novae phenomenon \cite[]{2016MNRAS.462.4435T,2017A&A...604A..48S}.

CN Cha serves as an example of slow symbiotic novae, exhibiting an extended, low-luminosity nova event that represents the lowest recorded luminosity \cite[]{2020AJ....160..125L}. This system comprises a white dwarf ($\sim0.6\mathrm{M_\odot}$) and a Mira-type donor, which played a pivotal role in contributing to the luminosity from 1963 until 2013. The estimated accretion rate for this system is approximately $\sim10^{-8}\mathrm{M_\odot}$ \cite[]{2023ApJ...951..128K}. Following the onset of the outburst phase in 2013, CN Cha maintained relatively constant luminosity for around three years before transitioning into the dimming phase, experiencing a decrease of 1.4 magnitudes per year \cite[]{2020AJ....160..125L,2023ApJ...951..128K}. 
For low mass WDs, of order $M_{\rm WD}\sim0.6-0.7\mathrm{M_\odot}$ the acceleration of the wind can often be too weak to emit optically thick winds, thus the nova envelope evolves very slowly, resulting in a flat optical peak during the outburst \cite[]{2009ApJ...699.1293K,2011ApJ...743..157K} as seen in CN Cha. Prolonged maxima have been observed in novae in CVs as well \cite[]{2010AJ....140...34S,2020A&A...635A.115M} and models show them to be low mass WDs as well, of order $<0.7\mathrm{M_\odot}$ accreting at low rates of order $\leq10^{-10}\mathrm{M_\odot} {\rm yr}^{-1}$ \cite[]{2022MNRAS.515.1404H}. 

 However, while in CVs it expected that novae that occur in a certain system will be relatively similar to each other, even if there are thousands of years between two consecutive novae, because the changes in the system (accretion rate etc) are slow. We cannot assume such consistency in novae that occur in symbiotic systems, because the system evolves on a shorter timescale and the accretion rate could be very different from one cycle to the next. This explains why AG Peg has demonstrated stark different behaviors from one eruption to the next. Therefore, while CN Cha demonstrates features of prolonged eruptions, typical of slow accretion, we cannot deduce that this will remain as-is and it is possible that in the future it will exhibit an eruption with entirely different features.

\section{Conclusions}\label{sec:conclusions}
The intention of this work was to study the characteristic behaviour of novae in symbiotic systems, that accrete matter from the wind of their donor via the BHL method. We have carried out a series of self-consistent nova simulations in symbiotic systems hosting an AGB donor of $1.0\mathrm{M_\odot}$ by changing the mass of the WD and the binary separation. We have taken into account the AML by MB, GR and drag,and how these physical mechanisms can affect the rate at which the mass is accreted onto the surface of the WD.  

We examined the behavior on a cyclic scale, single cycles, as well as the long term evolution of the system and compared our findings with novae in CVs. We found that the outcome of a single nova in a symbiotic system shows features similar to those of the outcome of a nova in a CV system, for a given WD mass and accretion rate. However, the long-term behavior is entirely different.

In CVs, the long-term behavior of the separation is to slowly and secularly decrease \cite[]{2020NatAs...4..886H,2021MNRAS.505.3260H}. We find that in symbiotic systems, the separation decreases as well, but on a much shorter time-scale, due to the substantial effect of drag in these systems. Additionally, in symbiotic systems, the orbital period may change abruptly depending on the wind rate of the AGB donor, whereas in CVs, there is always a decrease because the change in mass of the system is very slow with respect to the change in separation.We found that both the change in separation and the the change in donor mass determine whether the orbital period will increase or decrease, depending on which change is more substantial. A more substantial decrease in \textit{separation} will lead to a \textit{decrease} in the orbital period, whereas if the \textit{mass} decrease is more substantial, it will lead to an \textit{increase} in the orbital period. We find that smaller separations and more massive WDs lead to higher accretion rates, which is the regime where we find RN and therefore WDs that can grow. If the evolutionary time were longer, we would see a general trend of increasing accretion rates due to a decrease in separation. However, we are limited by the lifetime of the donor.

The accretion rate in CVs is not constant either, (see \citealt{2020NatAs...4..886H,2021MNRAS.505.3260H}). However, the changes in accretion rate in CVs occur in a manner distinct from what is obtained here. Here we find the accretion rate to exhibit periods of high and low values that are influenced by the wind rate. This causes abrupt changes in the mass retention efficiency, whereas in CVs it would be a slow and unnoticeable effect. \textit{If the donor star would somehow be able to have sufficient mass, any model featuring a positive $\Delta M_{WD}$ might potentially be considered as a viable candidate for a progenitor of SNIa}. 

There were some cycles with no mass ejection, supporting the possibility of weak TNR taking place on the surface of the WD. We find that, non-ejective eruptions occurs for low mass WDs at a lower $\dot M_{\mathrm{acc}}$ than they do for high mass WDs. This can mean that there is the possibility that such a system can have a positive $\eta$, i.e., net mass growth, and as the WD grows, the $\eta$ may evolve to become negative. However, the existence of parameter combinations that enable the WD to gain mass, suggests
that if such combinations can be achieved and maintained for more massive WDs,
they could eventually become progenitors of SNIa.

\section*{Data availability}
The data co-related with this work will be shared on reasonable request to the corresponding author.

\section*{Acknowledgments}
We thank the referee, Elias Aydi, for very helpful comments that helped improving the paper.
We thank Dina Prialnik for very helpful comments. We acknowledge support from the Ariel University Research and Development authority. We acknowledge the Ariel HPC Center at the Ariel University for providing computing resources.
YH acknowledges the support of the European Union's Horizon 2020 research and innovation program under grant agreement No 865932-ERC-SNeX.

\bibliographystyle{mnras}
\bibliography{main}
\end{document}